\documentclass[12pt, draftclsnofoot, onecolumn, journal]{IEEEtran}

\usepackage{booktabs}
\usepackage{amssymb}
\usepackage{color}
\usepackage[noadjust]{cite}
\usepackage{algorithm}
\usepackage{mathrsfs}
\usepackage{graphicx}
\usepackage{multicol}
\usepackage{indentfirst}
\usepackage{bm}
\usepackage{amsmath}
\usepackage{extarrows}
\usepackage{mathtools}

\newtheorem{remark}{\bf{Remark}}

\newtheorem{proposition}{\bf{Proposition}}

\usepackage{float}
\usepackage{multirow}
\usepackage{changepage}
\usepackage{amsfonts}
\usepackage{stfloats}
\usepackage{array}
\usepackage{titletoc}
\usepackage{algorithm}
\usepackage{algorithmic}

\ifCLASSOPTIONcompsoc
\usepackage[caption=false,font=normalsize,labelfont=sf,textfont=sf]{subfig}
\else
\usepackage[caption=false,font=footnotesize]{subfig}
\fi

\begin{document}

\title{Asynchronous RIS-assisted Localization: A Comprehensive Analysis of Fundamental Limits}
% \title{Capacity Results for Range-Limited SISO and MISO Dimmable VLC Channels}
\author{
    Ziyi Gong, \IEEEmembership{Student Member, IEEE,} Liang Wu, \IEEEmembership{Senior Member, IEEE,} Zaichen Zhang, \IEEEmembership{Senior Member, IEEE,} Jian Dang, \IEEEmembership{Senior Member, IEEE,} Yongpeng Wu, \IEEEmembership{Senior Member, IEEE,} and Jiangzhou Wang, \IEEEmembership{Fellow, IEEE}
    \thanks{
        % This work was supported in part by NSFC projects
        % under Grants 61960206005, 61871111, 62171127, 61971136, 61803211, and 62101127, in part by Postgraduate Research \& Practice Innovation Program of Jiangsu Province under Grant KYCX21\_0106, and in part by Research Fund of National Mobile Communications Research Laboratory.   
        % \emph{(Corresponding authors: Liang Wu; Zaichen Zhang.)}

        Ziyi Gong, Liang Wu, Zaichen Zhang and Jian Dang are with the National Mobile Communications Research Laboratory, Frontiers Science Center for Mobile Information Communication and Security, Southeast University, Nanjing 210096, China. Liang Wu, Zaichen Zhang and Jian Dang are also with Purple Mountain Laboratories, Nanjing 211111, China (email: \{ziyigong, wuliang, zczhang, dangjian\}@seu.edu.cn).

        Yongpeng Wu is with the Departent of Electronics Engineering, Shanghai Jiao Tong University, Minhang 200240, China ( email: yongpeng.wu@sjtu.edu.cn).

        Jiangzhou Wang is with the School of Engineering, University of Kent, Canterbury CT2 7NT, U.K. Email: (e-mail: j.z.wang@kent.ac.uk).
    }
}

\IEEEaftertitletext{\vspace{-1.5\baselineskip}}

\maketitle

\begin{abstract}
    The reconfigurable intelligent surface (RIS) has drawn considerable attention for its ability to enhance the performance of not only the wireless communication but also the indoor localization with low-cost. 
    This paper investigates the performance limits of the RIS-based near-field localization in the asynchronous scenario, and analyzes the impact of each part of the cascaded channel on the localization performance. The Fisher information matrix (FIM) and the position error bound (PEB) are derived. Besides, we also derive the equivalent Fisher information (EFI) for the position-related intermediate parameters. Enabled by the derived EFI, we verify that both the ranging and bearing information of the user can be obtained when the near-field model is considered for the RIS-User equipment (UE) part of the channel, while only the direction of the UE can be inferred in the far-field scenario. \textcolor{blue}{This result is well known in the scenario that the curvature of arrival (COA) is directly sensed by the traditional active large-scale array, and we prove that it still holds when the COA is sensed passively by the large RIS.}  For the base station (BS)-RIS part of the channel, we reveal that this part of the channel determines the type of the gain provided by the BS antenna array. \textcolor{blue}{Besides, in the single-carrier, single snapshot case, it requires both the BS-RIS and the RIS-UE part of the channel works in the near-field scenario to localize the UE.}
    % The multiple antennas at the BS can provide independent spatial gain in the near-field scenario, while only the power gain is available when the far-field model is considered. 
    We also show that the well-known focusing control scheme for RIS, which maximizes the received SNR, is not always a good choice and may degrade the localization performance in the asynchronous scenario. 
    The simulation results validate the analytic work. The impact of the focusing control scheme on the PEB performances under synchronous and asynchronous conditions is also investigated.    
    % Through the numerical results, the theoretic analyses of this paper are verified, and the different influences of the focusing control scheme on the PEB performances under synchronous and asynchronous conditions are also investigated. 
\end{abstract}

\section{Introduction}
High accuracy indoor positioning enabled by the wireless communication system has recently attracted \textcolor{blue}{considerable attention} and become a key requirement for the next generation cellular networks \cite{survey}. Exploiting high frequency, which has been regarded as a development direction in future wireless communication systems, makes it possible to deploy large-scale arrays \cite{2017Xiao}. In traditional massive multiple-input multiple-output (MIMO) systems, active large-scale arrays are equipped at the base stations (BSs). It has been shown that the active large-scale antenna array can significantly improve the localization accuracy  \textcolor{blue}{by providing high precision for direction of arrival (DOA) estimation} \cite{2017Nil}, mitigating the multi-path effect in positioning \cite{2019yunlong} and eliminating the pseudo-peaks in the MUSIC spectrum \cite{2021Gong}. The aforementioned works were all based on the far-field assumption. While when the array size is large enough, the near-field effect sometimes cannot be ignored. In these circumstances, the far-field planar wavefront assumption is untenable, and the spherical wavefront model must be considered. Although the spherical wavefront model usually makes the computational complexity increase (since the steering vector of the array becomes complicated), the curvature of arrival (COA) provided by the spherical wavefront could bring extra information about the user position, which can be beneficial especially when the system is lack of synchronization \cite{2019Zhang}, \cite{2021Guerra}. 
However, when the size of the active antenna array increases, both the cost and the power consumptions will also dramatically increase. Besides, most of the aforementioned advantages are based on the condition that the line-of-sight (LOS) path exists, while the active antenna arrays equipped at the BS usually cannot be flexibly deployed or smartly control the wireless propagation environment, which makes it hard to localize the user in the non-line-of-sight (NLOS) scenario.

Reconfigurable intelligent surface (RIS), which is regarded as a passive reflecting array with reconfigurable reflecting coefficients, has shown its potential to enhance the performance of not only the wireless communication but also the user localization with barely no extra power consumption \cite{2019Huang}–\cite{2022Pan}. RIS can be flexibly deployed in an appropriate position. Hence by adjusting the phase shifts, RIS can be used to artificially establish controllable NLOS links to cover the area where the LOS path from the user equipment (UE) to BS is obstructed \cite{2020Zhang}. Besides, the size of the RIS is usually large. This can be ascribed to a twofold reason. Firstly, the received signal-to-noise ratio (SNR) increases quadratically with the number of RIS elements \cite{2022magazine}, thus the size of the RIS must be large enough to establish a reliable link; secondly, unlike the traditional active large-scale array, we can deploy the large-scale RIS with low cost due to its passive characteristic. As a result, the near-field model must be applied when the large-scale RIS is employed. Therefore, theoretically investigating  the role of the large-scale RIS in the near-field wireless localization is imperative. In \cite{2021Guidi}, the effect of the RIS in radio positioning was investigated from the electromagnetic perspective of view, where the RIS-based positioning was regarded as a suitable mix of processing at electromagnetic and signal levels. The Fisher information theory \cite{EFIM}, \cite{Fundamentals Statistical Signal} is another important tool to evaluate the performance limits of the RIS-based localization. Based on the Fisher information theory, the corresponding Cramér Rao lower bounds (CRLBs) for the estimation of the user position or the related intermediate parameters can be obtained. In \cite{2021LiuICC}, the CRLBs for the intermediate parameters and the user position has been derived in the synchronous scenario when the LOS path was obstructed and the localization system was assisted by a single RIS. The phase design of the RIS has also been proposed which aimed to minimize the derived CRLB. In \cite{2021Liu}, \cite{2021Keykhosravi} and \cite{2022magazine}, the CRLBs for the situation that both the LOS path and the NLOS path established by a RIS exist have been derived in both synchronous and asynchronous scenarios. \cite{2022Ma} and \cite{2020Wymeersch} studied the CRLBs for the multiple RIS scenario. However, all the aforementioned works were based on the far-field assumption. When the near-field spherical wavefront was considered,  the CRLB for the user position has been investigated in the ideal synchronous condition \cite{2022Luan}, \cite{2022WangZY}.  In \cite{2021Elzanaty},  the CRLBs for not only the estimation of the user position but also the orientation and the position-related parameters were derived when the LOS path existed. However, the path-loss in \cite{2021Elzanaty} was assumed to accurately obey the free-space propagation rule, which was not realistic. Besides, the synchronization mismatch was \textcolor{blue}{not included in the measurement vector, which means that the analyses in \cite{2021Elzanaty} were actually based on the condition that the synchronization mismatch is known in advance.} The path-loss and the phase offset caused by the synchronization mismatch were treated as two independent unknown parameters in \cite{2021Rahal} and \cite{2021Abu-Shaban}. However, only the single-input and single-output (SISO) scenario was considered, and the impact of the multiple antennas at the BS was not investigated. Besides, in \cite{2022WangZY} and \cite{2021Elzanaty}, maximizing the received SNR was adopted as the phase design criterion for the RIS. This is not always a good choice in the asynchronous NLOS scenario, which will be discussed in this paper. \textcolor{blue}{In the above works which focused on the near-field scenario \cite{2022Luan}–\cite{2021Abu-Shaban}, the analysis of the multi-paths effect in the RIS link is missing, and most of the works \cite{2022Luan}, \cite{2021Elzanaty}–\cite{2021Abu-Shaban} did not take the amplitude differences across the RIS into account, which is not accurate.} More importantly, most of the aforementioned works did not systematically discuss how the channel characteristics, or in other words the channel-related system configurations, influence the CRLB in the RIS-aided near-field localization. Besides, the comparisons between the near-field and far-field effect in not only the RIS-UE part but also the BS-RIS part of the channel have not been presented.

In this paper, we investigate the performance limits of the single RIS assisted multiple-input and single-output (MISO) localization when the LOS path is obstructed and the near-field spherical wavefront model is considered in the asynchronous scenario. In these circumstances, the position information is only provided by the COA at the RIS. Furthermore, unlike the traditional localization based on the active large-scale antenna array, the channel of the RIS-based localization is a cascaded channel and consists of three parts. To the best of our knowledge, this paper is the first work that comprehensively analyzes the impact of each part of the cascaded channel on the performance limits of the localization by employing the equivalent Fisher information (EFI) theory \cite{EFIM}. The main contributions of this paper are summarized as follows:
\begin{itemize}
    \item In the asynchronous condition, we derive the corresponding Fisher information matrix (FIM), the position error bound (PEB), and the EFI for the position-related intermediate parameters. \textcolor{blue}{We adopt a more precise model that takes into account the amplitude differences across both the RIS and the BS array. The multi-paths effect between the BS and the RIS is also considered. Besides, we reveal that the information carried by the multi-paths between the BS and the RIS is also able to help improve the localization performance.} 
    \item Based on the derived EFI,  we verify that
    it is theoretically possible to localize the user with a single RIS when the near-field spherical wavefront is considered in the RIS-UE part of the channel because both the ranging and the bearing information can be obtained. \textcolor{blue}{However, the equivalent Fisher information for ranging parameter will tend to $0$ when the distance between the UE and the RIS tends to infinity, which means that only the bering information can be effectively inferred in the far-field scenario. The above results are well-known in the scenario that the COA is sensed by the traditional active large-scale array, and we prove that they still hold when the COA is sensed passively by the large RIS.} 
    \item When the near-field model is considered for the BS-RIS part of the channel, we reveal that the multiple antennas at the BS can provide independent spatial gain for the localization performance, while only the power gain can be achieved when this part of the channel works in the far-field scenario. Besides, \textcolor{blue}{we show that in the single-carrier, single snapshot case, it requires both the BS-RIS part and the RIS-UE part of the channel work in the near-field scenario to localize the UE. The role of the near-field effect in the BS-RIS part of the channel is to provide sufficient degrees of freedom to extract the COA information passively sensed by the RIS.}
    \item We show that unlike the synchronous scenario, the well-known focusing control scheme, which maximizes the received SNR, is not always a good choice in the RIS assisted asynchronous scenario because it may degrade the localization performance.
    % We also analyze the different influences of the focusing control scheme on the PEB performances under synchronous and asynchronous conditions through numerical results.
\end{itemize}

The rest of the paper is organized as follows. Section \ref{sec:signal model} presents the localization scenario and the signal model. In Section \ref{sec:results}, the results of the PEB and the EFI of the intermediate parameters are provided. The impact of each part of the cascaded channel on the localization performance is analyzed in detail in Section \ref{sec:IV}. Numerical results and discussions are provided in Section \ref{sec:numerical results} and conclusions are drawn in Section \ref{sec:conclusion}.

\emph{Notations:}   Upper and lower case bold symbols represent the matrices or column vectors. $(\cdot)^\top$, $(\cdot)^\mathrm{H}$ and $(\cdot)^{-1}$ denote the the transpose, the conjugate transpose (Hermitian) and the inverse of a matrix, respectively. $\Re\{ \cdot\}$ and $\Im\{ \cdot\}$ are the real and imaginary operators, $(\cdot)^*$ denotes the conjugate operator, $\mathbb{E}\{\cdot\}$ denotes the expectation operator, $\mathbb{D}\{\cdot\}$ denotes the variance operator, and $\circledast$ denotes the Hadamard product. $\mathrm{diag}\{\mathbf{a}\}$ represents a diagonal matrix with the elements of vector $\mathbf{a}$ on the main diagonal, $||\mathbf{a}||$ denotes the $\ell_2$-norm of the vector $\mathbf{a}$. $[\mathbf{A}]_{i,j}$ represents the $(i,j)$-th entry of matrix $\mathbf{A}$, $[\mathbf{A}]_{(r_1:r_2,c_1:c_2)}$ denotes the submatrix of matrix $\mathbf{A}$ composed of rows from $r_1$ to $r_2$ and columns from $c_1$ to $c_2$. $\mathrm{tr}(\mathbf{A})$ denotes the trace of matrix $\mathbf{A}$. $\mathbf{A}\succeq \mathbf{B}$ or $\mathbf{B}\preceq\mathbf{A}$ means that $\mathbf{A}-\mathbf{B}$ is positive semi-definite.  $\mathbf{1}_N$ denotes the $N$ dimensional all one vector. $\setminus$ is the set subtraction operator.

\section{Signal Model} \label{sec:signal model}
\begin{figure}[t]
    \centering  
    \color{blue}
    \includegraphics[width=8cm]{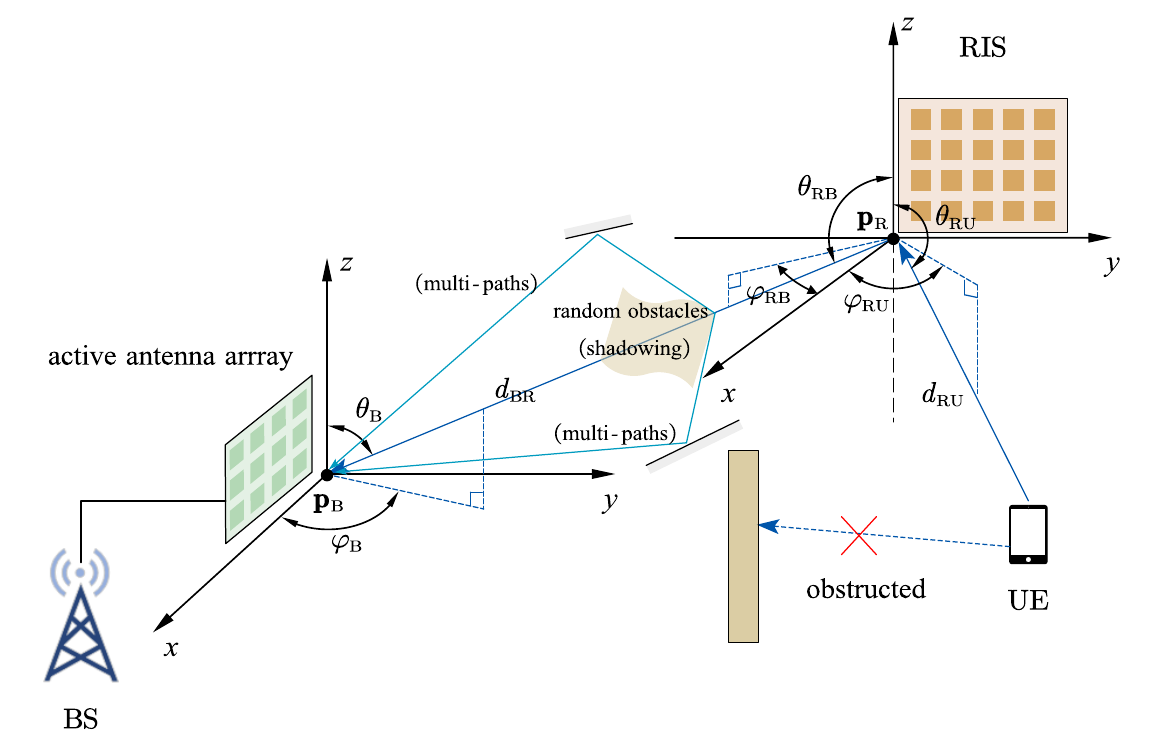}
    \caption{The geometry of the considered 3D RIS-assisted uplink localization scenario.}
    \label{3Dscneario}
\end{figure}

In this section, we present the three dimensional (3D) RIS-assisted localization scenario and the signal model that will be used in the following Fisher information analysis. This paper considers a RIS-assisted millimeter wave (mmWave)  asynchronous orthogonal frequency-division multiplexing  (OFDM) wireless system with $N$ sub-carriers \cite{OFDM1}, \cite{OFDM2}. As shown in Fig. \ref{3Dscneario}, the BS, which is equipped with a uniform rectangular array (URA) with $N_\mathrm{B}$ antennas and the reference point located in $\mathbf{p}_\mathrm{B}=[x_\mathrm{B}, y_\mathrm{B}, z_\mathrm{B}]^\top$, performs the localization for a single-antenna UE, located in $\mathbf{p}_\mathrm{U}=[x_\mathrm{U}, y_\mathrm{U}, z_\mathrm{U}]^\top$. The LOS path between the UE and the BS is obstructed. Therefore, in the uplink transmission, the BS receives the signal transmitted by the UE through the uncontrollable multi-paths reflected by the environment and the controllable path established by the RIS.  The RIS is assumed to be a passive URA with $N_\mathrm{R}$ elements. The reference point of the RIS is located in $\mathbf{p}_\mathrm{R}=[x_\mathrm{R}, y_\mathrm{R}, z_\mathrm{R}]^\top$. The positions of the RIS and the BS are assumed to be known. \textcolor{blue}{We also assume that there exists random obstacles between the BS and RIS, which cause shadowing effect and introduce multi-paths in this part of the channel.} Then the received signal at the BS for the $n$-th sub-carrier in the $t$-th time slot can be expressed as
\begin{align}
    \label{receive_sig}
    \color{blue}
    % \mathbf{\tilde{y}}_{n,t}&=x_{n,t} e^{-j2\pi f_n\xi} \biggl( \alpha \mathbf{H}_{\mathrm{BR},n}\mathbf{\Phi }_t\mathbf{h}_{\mathrm{RU},n}+ {\sum_i{\beta _i\mathbf{h}_{\mathrm{BU},n}^{\left( i \right)}}} \biggr)   \nonumber\\ 
    % &\mathrel{\phantom{=}}+{\mathbf{\tilde{w}}}_{n,t}. \;
    \bar{\mathbf{y}}_{n,t}=x_{n,t}e^{-j2\pi f_n\xi}\biggl( \alpha \mathbf{H}_{\mathrm{BR},n}\mathbf{\Phi }_t\mathbf{\Upsilon }_{\mathrm{RU},n}\mathring{\mathbf{h}}_{\mathrm{RU},n}+\sum_i{\beta _i\mathbf{h}_{\mathrm{BU},n}^{\left( i \right)}} \biggr) +\bar{\mathbf{w}}_{n,t}.\;
\end{align}
where $\xi$ is the unknown phase shift caused by the synchronization mismatch between the BS and the UE\footnote{\textcolor{blue}{The synchronization mismatch usually cannot be ignored even after a common uplink synchronization procedure. That is because the synchronization accuracy requirements are often in the microsecond range in most of the cellular networks, which cannot meet the requirement of high accuracy positioning.}}; $f_n$ is the frequency of the $n$-th sub-carrier; $x_{n,t}$ denotes the transmitted symbol. \textcolor{blue}{$\alpha$ represents the attenuation of the RIS link caused by the shadowing effect \cite{wireless communication}.} Unlike many existing works \cite{2022Dardari}, \cite{2021Elzanaty}–\cite{2022Proc.IEEE} assuming the free-space path-loss, here we assume that $\alpha$ is an independent unknown parameter, which is more realistic since it is hard to determine the precise relationship between the propagation distance and the path-loss in the practical channel 
% This assumption is more realistic since in the practical channel, due to the shadowing effect caused by the random obstacles, it is hard to determine the precise relationship between the propagation distance and the path-loss 
\cite{fundmental wireless communication}. The vector ${\mathbf{\bar{w}}}_{n,t}$ is the Gaussian white noise vector with
variance $\sigma^2$. The matrix $\mathbf{\Phi}_t=\mathrm{diag}\{ \boldsymbol{\phi}_t\}$ represents the phase shift induced by the RIS at time $t$, where $\boldsymbol{\phi}_t\in \mathbb{C}^{N_\mathrm{R}\times 1}$ is the dynamic reflection coefficient vector\footnote{\textcolor{blue}{Here the frequency flat narrowband model is considered for RIS coefficients. To highlight the near-field effect, we mainly focus on the mmWave system in this work, where a small fractional bandwidth could lead to a quite wide absolute bandwidth.}} \cite{2022Lin}, \cite{2021Elzanaty}.   
The vector \textcolor{blue}{$\mathring{\mathbf{h}}_{\mathrm{RU},n}\in \mathbb{C}^{N_\mathrm{R}\times 1}$} indicates the channel phase response between the RIS and the UE, with the $r$-th ($r=1,\cdots,N_\mathrm{R}$) element given by $\left[ \mathbf{h}_{\mathrm{RU},n} \right] _r=e^{-j2\pi f_nd_{r\mathrm{U}}/c}$,
% \begin{align}
% \left[ \mathbf{h}_{\mathrm{RU},n} \right] _r=e^{-j2\pi f_nd_{r\mathrm{U}}/c}, 
% \end{align}
where $d_{r\mathrm{U}}$ is the distance between the UE and the $r$-th element of the RIS, $c$ is the speed of light.
We consider both the exactly spherical wavefront model and the planar wavefront model. The former one reveals the near-field effect caused by the large size of the RIS \cite{2021Elzanaty}, \cite{2021Podkurkov} and the latter one represents the traditional far-field scenario. Let $\mathbf{p}_{r}$ denote the position of the $r$-th element on the RIS. Define $    \mathbf{p}_{\mathrm{R},r}=[x_{\mathrm{R},r}, y_{\mathrm{R},r}, z_{\mathrm{R},r}]^\top\triangleq \mathbf{p}_{r}- \mathbf{p}_\mathrm{R}$
which indicates coordinate of the $r$-th RIS element; $d_{\mathrm{RU}}\triangleq\| \mathbf{p}_{\mathrm{U}}-\mathbf{p}_{\mathrm{R}}\|$. Define $\theta _{\mathrm{RU}}$ and $\varphi _{\mathrm{RU}}$ as the elevation and azimuth angles of arrival at the RIS, respectively, as shown in Fig. \ref{3Dscneario}. Then we have
\begin{subequations} 
    \label{pRandpU}
    \begin{align}
        x_{\mathrm{U}}&=x_{\mathrm{R}}+d_{\mathrm{RU}}\sin  \theta _{\mathrm{RU}}  \cos  \varphi _{\mathrm{RU}} ,
        \\
        y_{\mathrm{U}}&=y_{\mathrm{R}}+d_{\mathrm{RU}}\sin  \theta _{\mathrm{RU}} \sin  \varphi _{\mathrm{RU}},
        \\
        z_{\mathrm{U}}&=z_{\mathrm{R}}+d_{\mathrm{RU}}\cos  \theta _{\mathrm{RU}} .
        \end{align}
\end{subequations}

For the case adopting the exactly spherical wavefront model, by combining (\refeq{pRandpU}), it can be derived that \cite{2021Elzanaty}
\begin{align}
    \label{dru_near}
    d_{r\mathrm{U}} = \| \mathbf{p}_{\mathrm{U}}-\mathbf{p}_{r}  \|=  
    \sqrt{\rho _{\mathrm{R},r}^{2}+d_{\mathrm{RU}}^{2}+2d_{\mathrm{RU}}\Gamma _{r\mathrm{U}}}, 
\end{align} 
where $\rho_{\mathrm{R},r}=\| \mathbf{p}_{\mathrm{R},r} \|$ denotes the distance between the $r$-th element and the reference point of the RIS;
%The following relationship holds
% \begin{align}
%     \mathbf{p}_{r}= \mathbf{p}_\mathrm{R}+\mathbf{p}_{\mathrm{R},r}.
% \end{align}
$\Gamma_{r\mathrm{U}}$ is given by 
\begin{align}
    \label{Gamma_rU}
    % \Gamma_{r\mathrm{U}}&= 
    % -x_{\mathrm{R},r}\sin \theta _{\mathrm{RU}}\cos \varphi _{\mathrm{RU}}-y_{\mathrm{R},r}\sin \theta _{\mathrm{RU}}\sin \varphi _{\mathrm{RU}} \nonumber \\
    % &\mathrel{\phantom{=}}-z_{\mathrm{R},r}\cos \theta _{\mathrm{RU}},
    \Gamma_{r\mathrm{U}}&= 
    -x_{\mathrm{R},r}\sin \theta _{\mathrm{RU}}\cos \varphi _{\mathrm{RU}}-y_{\mathrm{R},r}\sin \theta _{\mathrm{RU}}\sin \varphi _{\mathrm{RU}}-z_{\mathrm{R},r}\cos \theta _{\mathrm{RU}},
\end{align}

For the traditional far-field case that the planar wavefront model is adopted, eq. (\ref{dru_near}) is expanded to the first order term as \cite{fundmental wireless communication} 
\begin{align}
    \label{drU_far}
    d_{r\mathrm{U}} \simeq d_{\mathrm{RU}}+ \Gamma_{r\mathrm{U}}.
\end{align}

\textcolor{blue}{The matrix $\mathbf{\Upsilon }_{\mathrm{RU},n}=\mathrm{diag}\{ \boldsymbol{\gamma}_{\mathrm{RU},n} \}$ reflects the different amplitudes across the RIS in the near-field scenario. The $r$-th element of $\boldsymbol{\gamma}_{\mathrm{RU},n}\in \mathbb{R}^{N_\mathrm{R}\times 1}$ is $[\boldsymbol{\gamma}_{\mathrm{RU},n}]_{r}=\frac{\sqrt{2P_{\mathrm{t}}}\lambda _n}{4\pi d_{r\mathrm{U}}}$. Note that in the far-field scenario, $\mathbf{\Upsilon }_{\mathrm{RU},n}\to \frac{\sqrt{2P_{\mathrm{t}}}\lambda _n}{4\pi d_{\mathrm{RU}}}\mathbf{I}$.} 
% Compare (\refeq{dru_near}) and (\refeq{drU_far}), we can observe that the exactly spherical wavefront contains the position information of the UE, while from the planar wavefront only the direction information of the UE can be inferred.

The matrix $\mathbf{H}_{\mathrm{BR},n}\in\mathbb{C}^{N_\mathrm{B}\times N_\mathrm{R}}$ in (\refeq{receive_sig}) represents the channel phase response between the BS and the RIS, \textcolor{blue}{which is expressed as $\mathbf{H}_{\mathrm{BR},n}=\mathbf{H}_{\mathrm{BR},n}^{\left( \mathrm{d} \right)}+\mathbf{H}_{\mathrm{BR},n}^{\left( \mathrm{multi} \right)}$. The matrix $\mathbf{H}_{\mathrm{BR},n}^{\left( \mathrm{d} \right)}$ represents the channel with respect to the LOS path between the BS and the RIS.} Here we also consider both the near-field spherical wavefront and the far-field planar wavefront in this part of the channel. Then the $(b,r)$-th entry of  $\mathbf{H}^{\left( \mathrm{d} \right)}_{\mathrm{BR},n}$ in both cases can be expressed as  
\textcolor{blue}{
\begin{align}
   [ \mathbf{H}^{\left( \mathrm{d} \right)}_{\mathrm{BR},n} ] _{b,r}&= \frac{\lambda _n}{4\pi d_{br}}e^{-j2\pi f_nd_{br}/c}  &\text{(near-field)} \label{HBR_near}
    \\
    &\simeq \frac{\lambda _n}{4\pi d_{\mathrm{BR}}}e^{-j2\pi f_nd_{br}/c}  e^{-j2\pi f_nd_{\mathrm{BR}}/c} \left[ \mathbf{a}_{\mathrm{B}} \mathbf{a}_{\mathrm{RB}}^{\top} \right] _{b,r} \; &\text{(far-field)} \label{HBR_far}
\end{align}} 
where $d_{br}$ denotes the distance between the $b$-th antenna at the BS and the $r$-th element of the RIS; $d_\mathrm{BR}=\|\mathbf{p}_\mathrm{B}-\mathbf{p}_\mathrm{R}\|$; $\mathbf{a}_{\mathrm{B}}\in\mathbb{C}^{N_\mathrm{B}\times 1}$ and $\mathbf{a}_{\mathrm{RB}}\in\mathbb{C}^{N_\mathrm{R}\times 1}$ are the far-field array steering vectors at the BS and the RIS, respectively. The $b$-th element of  $\mathbf{a}_{\mathrm{B}}$ and the $r$-th element of  $\mathbf{a}_{\mathrm{RB}}$ are given by 
\begin{align}
    \label{aB_aRB}
    [\mathbf{a}_{\mathrm{B}}]_b=
    e^{-j2\pi f_n\Gamma _{\mathrm{B},b}} ,\quad    [\mathbf{a}_{\mathrm{RB}}]_r=
    e^{-j2\pi f_n\Gamma _{r\mathrm{B}}}, 
\end{align}
where 
\begin{align}
    % \Gamma _{\mathrm{B},b} &={ -}x_{\mathrm{B},b}\sin \theta _{\mathrm{B}}\cos \varphi _{\mathrm{B}}{-}y_{\mathrm{B},b}\sin \theta _{\mathrm{B}}\sin \varphi _{\mathrm{B}}{-}z_{\mathrm{B},b}\cos \theta _{\mathrm{B}}, \nonumber \\
    % \Gamma _{r\mathrm{B}}&=-x_{\mathrm{R},r}\sin \theta _{\mathrm{RB}}\cos \varphi _{\mathrm{RB}}-y_{\mathrm{R},r}\sin \theta _{\mathrm{RB}}\sin \varphi _{\mathrm{RB}}\nonumber\\
    % &\mathrel{\phantom{=}}-z_{\mathrm{R},r}\cos \theta _{\mathrm{RB}}.
    \Gamma _{\mathrm{B},b} &={ -}x_{\mathrm{B},b}\sin \theta _{\mathrm{B}}\cos \varphi _{\mathrm{B}}{-}y_{\mathrm{B},b}\sin \theta _{\mathrm{B}}\sin \varphi _{\mathrm{B}}{-}z_{\mathrm{B},b}\cos \theta _{\mathrm{B}}, \nonumber \\
    \Gamma _{r\mathrm{B}}&=-x_{\mathrm{R},r}\sin \theta _{\mathrm{RB}}\cos \varphi _{\mathrm{RB}}-y_{\mathrm{R},r}\sin \theta _{\mathrm{RB}}\sin \varphi _{\mathrm{RB}}-z_{\mathrm{R},r}\cos \theta _{\mathrm{RB}}.
\end{align}
Similar to the definition of $\mathbf{p}_{\mathrm{R},r}$, $\mathbf{p}_{\mathrm{B},b}\triangleq[x_{\mathrm{B},b}, y_{\mathrm{B},b}, z_{\mathrm{B},b}]^\top$ indicates the $b$-th antenna coordinate of the BS. The parameters $(\theta _\mathrm{RB}, \varphi _{\mathrm{RB}})$ and $(\theta _\mathrm{B}, \varphi _{\mathrm{B}})$ are the the elevation and azimuth angles of departure at the RIS and the angles of arrival at the BS, respectively.
It is noteworthy that although this part of the channel does not directly contain the information of the UE position, adopting whether the near-field spherical wavefront model or the far-field planar wavefront model will still significantly affect the CRLB performance in localization, which will be discussed in Section \ref{sec:BS-RIS} and verified in Section \ref{sec:numerical results}. \textcolor{blue}{The matrix $\mathbf{H}_{\mathrm{BR},n}^{\left( \mathrm{multi} \right)}$ represents channel respect to the multi-paths between the BS and the RIS. We adopt the independent, identically distributed (i.i.d.) Rayleigh fading model to characterize this part of the channel. Under this assumption, the entries of $\mathbf{H}_{\mathrm{BR},n}^{\left( \mathrm{multi} \right)}$ are i.i.d. circular symmetric complex Gaussian with variance $\sigma_\mathrm{H}^2$ \cite{fundmental wireless communication}.    }

The term $ x_{n,t}{\sum_i{\beta _i\mathbf{h}_{\mathrm{BU},n}^{\left( i \right)}}}$ in (\refeq{receive_sig}) represents the signal received from the multi-paths, where $\beta_i$ and $\mathbf{h}_{\mathrm{BU},n}^{\left( i \right)}$ are the power attenuation and the channel phase response of the $i$-th path, respectively. This term has no contribution to the localization of the UE since the positions of the reflectors are not known. In other words, this part of the channel is uncontrollable. Usually, these uncontrollable multi-paths reflected from other scatterers are ignored, thanks to the strong path degradation in mmWave propagation \cite{2022Gao}. For the case that the energy of the multi-paths cannot be ignored, one can adopt the signal components separation method introduced in \cite{2022Dardari} and \cite{2022magazine} to extract the RIS-reflected component by exploiting the time dimension. Therefore, to focus on the role that the RIS plays in the localization, in the following analysis, we only consider the received signal  from the path reflected by the RIS. Thus the received signal in (\refeq{receive_sig}) can be rewritten as
\textcolor{blue}{
\begin{align}   
    \mathbf{y}_{n,t}&=\alpha x_{n,t}e^{-j2\pi f_n\xi}\mathbf{H}_{\mathrm{BR},n}\mathbf{\Phi }_t\mathbf{\Upsilon }_{\mathrm{RU},n}\mathring{\mathbf{h}}_{\mathrm{RU},n}+\mathbf{w}_{n,t}, \label{receive_sig2}  \\
    &= \boldsymbol{\mu }_{n,t}+\tilde{\mathbf{w}}_{\mathrm{n},\mathrm{t}} \label{mu_nt} 
\end{align}
where 
\begin{align}   
    \label{mu and w}
    \boldsymbol{\mu }_{n,t}\triangleq \alpha x_{n,t}e^{-j2\pi f_n\xi}\mathbf{H}_{\mathrm{BR},n}^{(\mathrm{d})}\mathbf{\Phi }_t\mathbf{\Upsilon }_{\mathrm{RU},n}\mathring{\mathbf{h}}_{\mathrm{RU},n}, \quad
    \tilde{\mathbf{w}}_{\mathrm{n},\mathrm{t}}\triangleq\mathbf{H}_{\mathrm{BR},n}^{(\mathrm{multi})} \tilde{\mathbf{x}}_{n,t} +\mathbf{w}_{n,t}
\end{align}
with $\tilde{\mathbf{x}}_{n,t}=\alpha x_{n,t}e^{-j2\pi f_n\xi}\mathbf{\Phi }_t\mathbf{\Upsilon }_{\mathrm{RU},n}\mathring{\mathbf{h}}_{\mathrm{RU},n}$. The covariance matrix of the equivalent noise $ \tilde{\mathbf{w}}_{\mathrm{n},\mathrm{t}}$ is then given by
\begin{align}
    \label{Cwnt}
    \mathbf{C}_{\tilde{\mathbf{w}}_{n,t}}=\left( \left\| \tilde{\mathbf{x}}_{n,t} \right\| ^2\sigma _{\mathrm{H}}^{2}+\sigma ^2 \right) \mathbf{I}=( \alpha ^2|x_{n,t}|^2\left\| \boldsymbol{\gamma }_{\mathrm{RU},n} \right\| ^2\sigma _{\mathrm{H}}^{2}+\sigma ^2 ) \mathbf{I}.
\end{align}}
\vspace{-0.5cm}
% Note that according to the signal model given in (\refeq{receive_sig2}), the TOA information cannot be inferred due to the synchronization mismatch $\xi$ and the channel amplitude $\alpha$ is assumed to be an unknown independent parameter. However, the position information of the UE can be obtained from the spherical wavefront sensed by the RIS (namely COA).
\section{The Fundamental Limits} \label{sec:results}
In this section, based on the signal model (\refeq{receive_sig2}), we derive the FIM and the CRLB for not only the unknown UE position but also the intermediate parameters that are related to the position information. The EFI results for the intermediate parameters will enable the further analysis in the following section. This section focuses on the near-field scenario between the RIS and the UE. The EFI for the far-field scenario will be investigated and compared with that for the near-field scenario in Section \ref{sec:RIS-UE}. 
\subsection{The Position Error Bound}
According to the localization scenario and the signal model introduced in Section \ref{sec:signal model}, the unknown parameter vector that contains the UE position and the other nuisance parameters is given by $\mathbf{\Theta }=\left[ \alpha , c\xi , \mathbf{p}^{\top}_\mathrm{U} \right] ^{\top} \in \mathbb{R}^{5\times 1}$.
% \begin{align}
%     \mathbf{\Theta }=\left[ \alpha , c\xi , \mathbf{p}^{\top}_\mathrm{U} \right] ^{\top} \in \mathbb{R}^{5\times 1}.
% \end{align}
Suppose that the BS receives the signal in $T$ time slots. \textcolor{blue}{From (\refeq{mu and w}), we note that both the signal part $\boldsymbol{\mu}_{n,t}$ and the equivalent noise part $\tilde{\mathbf{w}}_{\mathrm{n},\mathrm{t}}$ are related to $\mathbf{\Theta}$.}  Therefore, the FIM for $\mathbf{\Theta }$ is calculated as \cite{2021Elzanaty}, \cite{Fundamentals Statistical Signal}
\begin{align}
    \color{blue}
    \label{FIMJp}
    % \mathbf{J}&=\sum_{t=1}^T{\sum_{n=1}^N{\frac{2}{\sigma ^2}}}\Re \left\{  \frac{\partial \boldsymbol{\mu }_{n,t} ^{\mathrm{H}}}{\partial \mathbf{\Theta }} \frac{\partial \boldsymbol{\mu }_{n,t}}{\partial \mathbf{\Theta }^{\top}} \right\} \nonumber
    % \\
    % &=\sum_{t=1}^T{\sum_{n=1}^N{\sum_{b=1}^{N_{\mathrm{B}}}{\mathbf{J}_{b,n,t}}}}
    \boldsymbol{\mathcal{J}}=\mathbf{J}_{\tilde{\mathbf{w}}}+\mathbf{J}_{\boldsymbol{\mu}}=\mathbf{J}_{\tilde{\mathbf{w}}}+\sum_{t=1}^T{\sum_{n=1}^N{\sum_{b=1}^{N_{\mathrm{B}}}{\mathbf{J}_{b,n,t}}}},
\end{align}   
where $\mathbf{J}_{b,n,t}$ represents the Fisher information for $\mathbf{\Theta}$ provided by the \emph{\textcolor{blue}{signal part}} received via the $n$-th sub-carrier at the $b$-th antenna during the time slot $t$, i.e., a single sample, and it is given by
\begin{align}
    \label{Jbnt1}
    \mathbf{J}_{b,n,t}=\frac{2}{\sigma ^2}\Re \left\{ \frac{\partial \mu _{b,n,t}^{*}}{\partial \mathbf{\Theta }}\frac{\partial \mu _{b,n,t}}{\partial \mathbf{\Theta }^{\top}} \right\} \in \mathbb{R}^{5\times5}
\end{align}
where 
% \begin{align}
%     \label{mu_bnt1}
%     \mu _{b,n,t}&\triangleq\left[ \boldsymbol{\mu }_{n,t} \right] _b \nonumber \\
%     &=\alpha x_{n,t}e^{-j2\pi f_n\xi}\mathbf{h}_{b\mathrm{R},n}^\top\mathbf{\Phi }_t\mathbf{h}_{\mathrm{RU},n} \nonumber \\
%     &=\alpha x_{n,t}e^{-j2\pi f_n\xi} \boldsymbol{\phi}_t^\top\mathbf{\tilde{h}}_{b,n}
% \end{align}
\begin{align}
    \color{blue}
    \label{mu_bnt1}
    \mu _{b,n,t}\triangleq \left[ \boldsymbol{\mu }_{n,t} \right] _b=\alpha x_{n,t}e^{-j2\pi f_n\xi}\mathbf{h}_{b\mathrm{R},n}^{\top}\mathbf{\Phi }_t\mathbf{\Upsilon }_{\mathrm{RU},n}\mathring{\mathbf{h}}_{\mathrm{RU},n}=\alpha x_{n,t}e^{-j2\pi f_n\xi}\boldsymbol{\phi }_{t}^{\top}\tilde{\mathbf{h}}_{b,n}
\end{align}
\textcolor{blue}{with $\mathbf{h}_{b\mathrm{R},n}\triangleq ([\mathbf{H}_{\mathrm{BR},n}^{\left( \mathrm{d} \right)}]_{b,:})^{\top}$ and }
\begin{align}
    \color{blue}
    \label{tilde hbn}
    \mathbf{\tilde{h}}_{b,n}\triangleq\mathbf{h}_{b\mathrm{R},n}\circledast\boldsymbol{\gamma }_{\mathrm{RU},n}\circledast \mathring{\mathbf{h}}_{\mathrm{RU},n}.
\end{align}
Note that the contribution of the inter-sample information for the parameters has been inherently included in \textcolor{blue}{$\mathbf{J}_{\boldsymbol{\mu}}$}. 
Define $\mathbf{d}_{\mathrm{RU}}\triangleq\left[ d_{1\mathrm{U}},...,d_{N_{\mathrm{R}}\mathrm{U}} \right] ^{\top}$; \textcolor{blue}{$\dot{\mathbf{h}}_{b,n}\triangleq\mathbf{h}_{b\mathrm{R},n}\circledast \mathring{\mathbf{h}}_{\mathrm{RU},n}\circledast \dot{\mathbf{d}}_{\mathrm{RU}}$ where $[ \dot{\mathbf{d}}_{\mathrm{RU}} ] _r=-\frac{1}{d_{r\mathrm{U}}^{2}}, (1\le r\le N_{\mathrm{R}})$;
\begin{subequations}
    \label{Dhp}
    \begin{align}    
        \mathbf{D}_{\mathring{\mathbf{h}}\mathbf{p}}\triangleq \frac{\partial \mathring{\mathbf{h}}_{\mathrm{RU},n}}{\partial \mathbf{p}_{\mathrm{U}}^{\top}}=\frac{-j2\pi f_n}{c}\mathrm{diag}\left\{ \tilde{\mathbf{h}}_{b,n} \right\} \frac{\partial \mathbf{d}_{\mathrm{RU}}}{\partial \mathbf{p}_{\mathrm{U}}^{\top}}; \\
        \mathbf{D}_{\boldsymbol{\gamma }\mathbf{p}}\triangleq \frac{\partial \boldsymbol{\gamma }_{\mathrm{RU},n}}{\partial \mathbf{p}_{\mathrm{U}}^{\top}}=\frac{\sqrt{2P_{\mathrm{t}}}\lambda _n}{4\pi}\mathrm{diag}\left\{ \dot{\mathbf{h}}_{b,n} \right\} \frac{\partial \mathbf{d}_{\mathrm{RU}}}{\partial \mathbf{p}_{\mathrm{U}}^{\top}};
        % &\mathbf{D}_{\mathbf{hp}}\triangleq \frac{\partial \boldsymbol{\gamma }_{\mathrm{RU},n}\circledast\mathring{\mathbf{h}}_{\mathrm{RU},n}}{\partial \mathbf{p}_{\mathrm{U}}^{\top}}=\left( \frac{-j2\pi f_n}{c}\mathrm{diag}\left\{ \tilde{\mathbf{h}}_{b,n} \right\} +\frac{\sqrt{2P_{\mathrm{t}}}\lambda _n}{4\pi}\mathrm{diag}\left\{ \dot{\mathbf{h}}_{b,n} \right\} \right) \frac{\partial \mathbf{d}_{\mathrm{RU}}}{\partial \mathbf{p}_{\mathrm{U}}^{\top}} \label{Dhp}
    \end{align}
\end{subequations}
$\mathbf{D}_{\mathbf{hp}}\triangleq \mathbf{D}_{\mathring{\mathbf{h}}\mathbf{p}}+\mathbf{D}_{\boldsymbol{\gamma }\mathbf{p}}$.}
Note that $\mathbf{J}_{b,n,t}$ is symmetric and 
\begin{align}
    \Re \left\{ \frac{\partial \mu _{b,n,t}^{*}}{\partial \alpha}\frac{\partial \mu _{b,n,t}}{\partial \left( c\xi \right)} \right\} =\Re \left\{ \frac{-j2\pi f_n\alpha |x_n|^2}{c}|\boldsymbol{\phi }_t^{\top}\mathbf{\tilde{h}}_{b,n} |^2 \right\} =0, \label{J12eq0} 
\end{align}
\textcolor{blue}{Combining (\refeq{Jbnt1})–(\refeq{J12eq0}), $\mathbf{J}_{b,n,t}$ can be calculated as:
\begin{align}
    \mathbf{J}_{b,n,t}=\left[ \begin{matrix}
        J_{\alpha \alpha}^{(b,n,t)}&		0&		( \mathbf{j}_{\alpha \mathbf{p}}^{(b,n,t)} ) ^{\top}\\
        0&		J_{\xi \xi}^{(b,n,t)}&		( \mathbf{j}_{\xi \mathbf{p}}^{(b,n,t)} ) ^{\top}\\
        \mathbf{j}_{\alpha \mathbf{p}}^{(b,n,t)}&		\mathbf{j}_{\xi \mathbf{p}}^{(b,n,t)}&		\mathbf{J}_{\mathbf{pp}}^{(b,n,t)}\\
    \end{matrix} \right] ,
\end{align}
where $\mathbf{j}_{\alpha \mathbf{p}}^{(b,n,t)}, \mathbf{j}_{\xi \mathbf{p}}^{(b,n,t)}\in \mathbb{R}^{3\times1}$, $\mathbf{J}_{\mathbf{pp}}\in \mathbb{R}^{3\times3}$,
\begin{align}
    \label{Jbnt2}
    J_{\alpha \alpha}^{(b,n,t)}&=\frac{2}{\sigma ^2}|x_{n,t}|^2|\boldsymbol{\phi}_t^{\top}\mathbf{\tilde{h}}_{b,n}|^2 ,  & J_{\xi \xi}^{(b,n,t)}&=\frac{8\pi ^2f_{n}^{2}\alpha ^2}{c^2\sigma^{2}}\left| x_{n,t} \right|^2| \boldsymbol{\phi }_{t}^{\top}\mathbf{\tilde{h}}_{b,n} |^2, \nonumber\\
    \mathbf{j}_{\alpha \mathbf{p}}^{(b,n,t)}&=\Re \biggl\{ \frac{2\alpha |x_{n,t}|^2}{\sigma ^2}\tilde{\mathbf{h}}_{b,n}^{\mathrm{H}}\boldsymbol{\phi }_{t}^{*}\boldsymbol{\phi }_{t}^{\top}\mathbf{D}_{\mathbf{hp}}\biggr\}^\top, \,&
    \mathbf{j}_{\xi \mathbf{p}}^{(b,n,t)}&=\Re \biggl\{ \frac{j4\pi f_n\alpha ^2|x_{n,t}|^2}{c\sigma ^2}\tilde{\mathbf{h}}_{b,n}^{\mathrm{H}}\boldsymbol{\phi }_{t}^{*}\boldsymbol{\phi }_{t}^{\top}\mathbf{D}_{\mathbf{hp}} \biggr\} ^{\top}, \nonumber\\
    \mathbf{J}_{\mathbf{pp}}^{(b,n,t)}&=\Re \biggl\{ \frac{2\alpha ^2|x_{n,t}|^2}{\sigma ^2}\mathbf{D}_{\mathbf{hp}}^{\mathrm{H}}\boldsymbol{\phi }_{t}^{*}\boldsymbol{\phi }_{t}^{\top}\mathbf{D}_{\mathbf{hp}} \biggr\}.
\end{align}}
\footnote{A generally more computational efficient (but less elegant) equivalent expression for the term $\mathrm{diag}\{ \mathbf{\tilde{h}}_{b,n} \} \frac{\partial \mathbf{d}_{\mathrm{RU}}}{\partial \mathbf{p}_{\mathrm{U}}^{\top}}$ and $\mathrm{diag}\{ \dot{\mathbf{h}}_{b,n} \} \frac{\partial \mathbf{d}_{\mathrm{RU}}}{\partial \mathbf{p}_{\mathrm{U}}^{\top}}$ in (\refeq{Dhp}) is $[\tilde{\mathbf{h}}_{b,n}, \tilde{\mathbf{h}}_{b,n}, \tilde{\mathbf{h}}_{b,n}] \circledast \frac{\partial \mathbf{d}_{\mathrm{RU}}}{\partial \mathbf{p}_{\mathrm{U}}^{\top}}$ and $[\dot{\mathbf{h}}_{b,n},\dot{\mathbf{h}}_{b,n},\dot{\mathbf{h}}_{b,n}]\circledast \frac{\partial \mathbf{d}_{\mathrm{RU}}}{\partial \mathbf{p}_{\mathrm{U}}^{\top}}$, respectively.}Note that $d_{r\mathrm{U}}$ is a direct function of $\mathbf{p}_\mathrm{U}$ when adopting the exactly spherical wavefront model, then the $r$-th row of the Jacobian matrix $\frac{\partial \mathbf{d}_{\mathrm{RU}}}{\partial \mathbf{p}_{\mathrm{U}}^{\top}}\in\mathbb{R}^{N_\mathrm{R}\times3}$ in (\refeq{Dhp}) in the near-field scenario (combining (\refeq{dru_near})) is given by
\begin{align}
    \label{DdRU_pU_near}
    \left[\frac{\partial \mathbf{d}_{\mathrm{RU}}}{\partial \mathbf{p}_{\mathrm{U}}^{\top}}\right]_{r,:}=\frac{\partial \|\mathbf{p}_\mathrm{U}-\mathbf{p}_r\|}{\partial \mathbf{p}_{\mathrm{U}}^{\top}}=\frac{\left( \mathbf{p}_{\mathrm{U}}-\mathbf{p}_r \right) ^{\top}}{d_{r\mathrm{U}}}.
\end{align}

\textcolor{blue}{The matrix $\mathbf{J}_{\tilde{\mathbf{w}}}$ represents the Fisher information provided by the \emph{equivalent noise part}. The $(i,j)$-th entry of $\mathbf{J}_{\tilde{\mathbf{w}}}$ ($1\le i,j \le 5$) is expressed as \cite{Fundamentals Statistical Signal}
\begin{align}
    [\mathbf{J}_{\tilde{\mathbf{w}}}]_{i,j}=\sum_{t=1}^T{\sum_{n=1}^N{\mathrm{tr}\left\{ \mathbf{C}_{\tilde{\mathbf{w}}_{n,t}}^{-1}\frac{\partial \mathbf{C}_{\tilde{\mathbf{w}}_{n,t}}}{\partial \left[ \mathbf{\Theta } \right] _i}\mathbf{C}_{\tilde{\mathbf{w}}_{n,t}}^{-1}\frac{\partial \mathbf{C}_{\tilde{\mathbf{w}}_{n,t}}}{\partial \left[ \mathbf{\Theta } \right] _j} \right\}}}.
\end{align}
Combining (\refeq{Cwnt}), we have $\frac{\partial \mathbf{C}_{\tilde{\mathbf{w}}_{n,t}}}{\partial \left( c\xi \right)}=\mathbf{0}$,
\begin{align}
    \frac{\partial \mathbf{C}_{\tilde{\mathbf{w}}_{n,t}}}{\partial \alpha}&=2\alpha |x_{n,t}|^2\left\| \boldsymbol{\gamma }_{\mathrm{RU},n} \right\| ^2\sigma _{\mathrm{H}}^{2}\mathbf{I}\triangleq {c}_{\alpha \left( n,t \right)}\sigma _{\mathrm{H}}^{2}\mathbf{I}, \label{c_alpha}\\
    \frac{\partial \mathbf{C}_{\tilde{\mathbf{w}}_{n,t}}}{\partial \left[ \mathbf{p}_{\mathrm{U}} \right] _k}&=2\alpha ^2\frac{\sqrt{2P_{\mathrm{t}}}\lambda _n}{4\pi}\left| x_{n,t} \right|^2\boldsymbol{\gamma }_{\mathrm{RU},n}^{\mathrm{T}}\left( \dot{\mathbf{d}}_{\mathrm{RU}}\circledast \frac{\partial \mathbf{d}_{\mathrm{RU}}}{\partial \left[ \mathbf{p}_{\mathrm{U}} \right] _k} \right) \sigma _{\mathrm{H}}^{2}\mathbf{I}\triangleq\left[ {\mathbf{c}}_{\mathbf{p}\left( n,t \right)} \right] _k\sigma _{\mathrm{H}}^{2}\mathbf{I}.
\end{align}
where $1\le k \le 3$, $\frac{\partial \mathbf{d}_{\mathrm{RU}}}{\partial [ \mathbf{p}_{\mathrm{U}} ] _k}=\bigl[\frac{\partial \mathbf{d}_{\mathrm{RU}}}{\partial \mathbf{p}_{\mathrm{U}}^{\top}}\bigr]_{:,k}$. Define ${\mathbf{c}}_{\mathbf{\Theta }\left( n,t \right)}=[ {c}_{\alpha \left( n,t \right)}, 0, {\mathbf{c}}^\top_{\mathbf{p}\left( n,t \right)} ] ^{\top}$, $\mathbf{J}_{\tilde{\mathbf{w}}}$ can be further calculated as
\begin{align}
    \label{Jw1}
    \mathbf{J}_{\tilde{\mathbf{w}}}=\sum_{t=1}^T{\sum_{n=1}^N{\frac{N_{\mathrm{B}}\sigma _{\mathrm{H}}^{4}}{\left( \alpha ^2|x_{n,t}|^2\left\| \boldsymbol{\gamma }_{\mathrm{RU},n} \right\| ^2\sigma _{\mathrm{H}}^{2}+\sigma ^2 \right) ^2}}}{\mathbf{c}}_{\mathbf{\Theta }\left( n,t \right)}{\mathbf{c}}_{\mathbf{\Theta }\left( n,t \right)}^{\top}.
\end{align}
}
Finally, the PEB is defined as\footnote{For any unbiased estimator $\hat{\mathbf{p}}$, $\mathrm{PEB}\le\sqrt{\mathbb{E}\{ \|\hat{\mathbf{p}}-\mathbf{p}\|^2\}}$.}  \cite{2021Elzanaty}, \cite{2021Abu-Shaban}
\begin{align}
    \mathrm{PEB}(\boldsymbol{\mathcal{J}}) = \sqrt{\mathrm{tr}([\boldsymbol{\mathcal{J}}^{-1}]_{3:5,3:5})}. 
\end{align}
\begin{remark}\label{remark1}
    Since 
    \begin{align}
        \color{blue}
        \mathbf{J}_{\boldsymbol{\mu}} - \mathbf{J}'_{\boldsymbol{\mu}} \triangleq \mathbf{J}_{\boldsymbol{\mu}}-\sum_{t,t\ne t'}{\sum_{n,n\ne n'}{\sum_{b,b\ne b'}{\mathbf{J}_{b,n,t}}}}=\mathbf{J}_{b',n',t'}\succeq \mathbf{0}, 
    \end{align}
    it can be verified that the CRLBs corresponding to $\mathbf{J}_{\boldsymbol{\mu}}$ and $\mathbf{J}'_{\boldsymbol{\mu}}$ satisfy $ \mathbf{J}_{\boldsymbol{\mu}}^{-1}\preceq (\mathbf{J}'_{\boldsymbol{\mu}})^{-1} $ \cite[Corollary 7.7.4]{Matrix Analysis}. In other words, adding a sample (with Fisher information $\mathbf{J}_{b',n',t'}$) will generally improve the localization performance \cite{Fundamentals Statistical Signal}. \textcolor{blue}{Similarly, the information carried by the multi-paths between the BS and the RIS, which corresponds to $\mathbf{J}_{\tilde{\mathbf{w}}}$, is also able to help improve the estimation performance. } Besides, we note that according to (\refeq{FIMJp}), \textcolor{blue}{$\mathbf{J}_{\boldsymbol{\mu}}$} intrinsically contains the information about $\mathbf{\Theta}$ provided by all the samples, and we can trade one of the space, bandwidth or time resource for the other two  while maintaining the same CRLB performance, which will be shown in Section \ref{sec:numerical results}.
\end{remark}

\subsection{EFI For the Intermediate Parameters}
In order to give more insights in the analysis of how the localization performance is influenced by the system configuration, we introduce the intermediate parameters, including the unknown distance and direction of the UE observed at the RIS, which is related to the UE position. The corresponding parameter vector is then defined as $\mathbf{\bar{\Theta}}=\left[ \alpha , c\xi , d_{\mathrm{RU}}, \varphi _{\mathrm{RU}}, \theta _{\mathrm{RU}} \right] ^{\top} \in \mathbb{R}^{5\times1}$.
% \begin{align}
%     \mathbf{\bar{\Theta}}=\left[ \alpha , c\xi , d_{\mathrm{RU}}, \varphi _{\mathrm{RU}}, \theta _{\mathrm{RU}} \right] ^{\top} \in \mathbb{R}^{5\times1}.
% \end{align}
The FIM of $ \mathbf{\bar{\Theta}}$ is given by \textcolor{blue}{$\bar{\boldsymbol{\mathcal{J}}}=\bar{\mathbf{J}}_{\tilde{\mathbf{w}}}+\bar{\mathbf{J}}_{\boldsymbol{\mu }}=\bar{\mathbf{J}}_{\tilde{\mathbf{w}}}+\sum_{t=1}^T{\sum_{n=1}^N{\sum_{b=1}^{N_{\mathrm{B}}}{\bar{\mathbf{J}}_{b,n,t}}}}.$
First we derive $\bar{\mathbf{J}}_{b,n,t}$. Define $\boldsymbol{\eta}=\left[ d_{\mathrm{RU}}, \varphi _{\mathrm{RU}}, \theta _{\mathrm{RU}} \right] ^{\top}$, 
\begin{align}
    \tilde{\mathbf{d}}_{\boldsymbol{\eta }_k}\triangleq \tilde{\mathbf{h}}_{b,n}\circledast \frac{\partial \mathbf{d}_{\mathrm{RU}}}{\partial [\boldsymbol{\eta }]_k}, \quad \dot{\mathbf{d}}_{\boldsymbol{\eta }_k}\triangleq \dot{\mathbf{h}}_{b,n}\circledast \frac{\partial \mathbf{d}_{\mathrm{RU}}}{\partial [\boldsymbol{\eta }]_k}.
\end{align}
where $1\le k \le 3$. 
Combining (\refeq{dru_near}) and (\refeq{Gamma_rU}), in the near-field scenario the $r$-th element of $\frac{\partial \mathbf{d}_{\mathrm{RU}}}{\partial [\boldsymbol{\eta }]_k}\in\mathbb{R}^{N_\mathrm{R}\times1}$ can be calculated as
\begin{subequations}
    \label{DdRU_eta_near}
    \begin{align}
        &\left[\frac{\partial \mathbf{d}_{\mathrm{RU}}}{\partial [\boldsymbol{\eta }]_1}\right]_{r}= \frac{\partial {d}_{r\mathrm{U}}}{\partial d_{\mathrm{RU}}} =\frac{1}{d_{r\mathrm{U}}}\left( d_{\mathrm{RU}}+\Gamma _{r\mathrm{U}} \right) ,  \label{DdRU_eta_near_1}\\
        &\left[\frac{\partial \mathbf{d}_{\mathrm{RU}}}{\partial [\boldsymbol{\eta }]_2}\right]_{r}=\frac{\partial {d}_{r\mathrm{U}}}{\partial \varphi _{\mathrm{RU}}} =\frac{d_{\mathrm{RU}}}{d_{r\mathrm{U}}}( x_{\mathrm{R},r}\sin \theta _{\mathrm{RU}}\sin \varphi _{\mathrm{RU}}-y_{\mathrm{R},r}\sin \theta _{\mathrm{RU}}\cos \varphi _{\mathrm{RU}} ) ,\\
        &\left[\frac{\partial \mathbf{d}_{\mathrm{RU}}}{\partial [\boldsymbol{\eta }]_3}\right]_{r}= \frac{\partial {d}_{r\mathrm{U}}}{\partial \theta _{\mathrm{RU}}}  =\frac{d_{\mathrm{RU}}}{d_{r\mathrm{U}}}( -x_{\mathrm{R},r}\cos \theta _{\mathrm{RU}}\cos \varphi _{\mathrm{RU}}-y_{\mathrm{R},r}\cos \theta _{\mathrm{RU}}\sin \varphi _{\mathrm{RU}}+z_{\mathrm{R},r}\sin \theta _{\mathrm{RU}} ),  
    \end{align}    
\end{subequations}
where $1\le r \le N_\mathrm{R}$.
Then $\mathbf{\bar{J}}_{b,n,t}$ can be expressed as
\begin{align}
    \label{Jbarbnt1}
    \bar{\mathbf{J}}_{b,n,t}=\left[ \begin{matrix}
        J_{\alpha \alpha}^{(b,n,t)}&		0&		(\mathbf{j}_{\alpha \boldsymbol{\eta }}^{(b,n,t)})^{\top}\\
        0&		J_{\xi \xi}^{(b,n,t)}&		(\mathbf{j}_{\xi \boldsymbol{\eta }}^{(b,n,t)})^{\top}\\
        \mathbf{j}_{\alpha \boldsymbol{\eta }}^{(b,n,t)}&		\mathbf{j}_{\xi \boldsymbol{\eta }}^{(b,n,t)}&		\mathbf{J}_{\boldsymbol{\eta \eta }}^{(b,n,t)}\\
    \end{matrix} \right] ,
\end{align}
where $ J_{\alpha \alpha}^{(b,n,t)}$ and $J_{\xi \xi}^{(b,n,t)}$ are defined in (\refeq{Jbnt2}), 
\begin{align}
    \label{Jbarbnt_ele}
    [\mathbf{j}_{\alpha \boldsymbol{\eta }}^{(b,n,t)}]_k&=\Re \biggl\{ \frac{2\alpha |x_{n,t}|^2}{\sigma ^2}\tilde{\mathbf{h}}_{b,n}^{\mathrm{H}}\boldsymbol{\phi }_{t}^{*}\boldsymbol{\phi }_{t}^{\top}\left( \frac{-j2\pi f_n}{c}\tilde{\mathbf{d}}_{\boldsymbol{\eta }_k}+\frac{\sqrt{2P_{\mathrm{t}}}\lambda _n}{4\pi}\dot{\mathbf{d}}_{\boldsymbol{\eta }_k} \right) \biggr\}, \nonumber \\
    [\mathbf{j}_{\xi \boldsymbol{\eta }}^{(b,n,t)}]_k&=\Re \biggl\{ \frac{j4\pi f_n\alpha ^2|x_{n,t}|^2}{c\sigma ^2}\tilde{\mathbf{h}}_{b,n}^{\mathrm{H}}\boldsymbol{\phi }_{t}^{*}\boldsymbol{\phi }_{t}^{\top}\left( \frac{-j2\pi f_n}{c}\tilde{\mathbf{d}}_{\boldsymbol{\eta }_k}+\frac{\sqrt{2P_{\mathrm{t}}}\lambda _n}{4\pi}\dot{\mathbf{d}}_{\boldsymbol{\eta }_k} \right) \biggr\}, \nonumber \\
    \left[ \mathbf{J}_{\boldsymbol{\eta \eta }}^{(b,n,t)} \right] _{k,l}&=\Re \biggl\{ \frac{8\pi ^2\alpha ^2f_{n}^{2}|x_{n,t}|^2}{c^2\sigma ^2}\tilde{\mathbf{d}}_{\boldsymbol{\eta }_k}^{\mathrm{H}}\boldsymbol{\phi }_{t}^{*}\boldsymbol{\phi }_{t}^{\top}\tilde{\mathbf{d}}_{\boldsymbol{\eta }_l} \biggr\} +\Re \left\{ \frac{\alpha ^2P_{\mathrm{t}}\lambda _{n}^{2}|x_{n,t}|^2}{4\pi ^2\sigma ^2}\dot{\mathbf{d}}_{\boldsymbol{\eta }_k}^{\mathrm{H}}\boldsymbol{\phi }_{t}^{*}\boldsymbol{\phi }_{t}^{\top}\dot{\mathbf{d}}_{\boldsymbol{\eta }_l} \right\} \nonumber\\
    &\mathrel{\phantom{=}}+\Re \left\{ \frac{j2\sqrt{2P_{\mathrm{t}}}\alpha ^2|x_{n,t}|^2}{\sigma ^2}\tilde{\mathbf{d}}_{\boldsymbol{\eta }_k}^{\mathrm{H}}\boldsymbol{\phi }_{t}^{*}\boldsymbol{\phi }_{t}^{\top}\dot{\mathbf{d}}_{\boldsymbol{\eta }_l} \right\},
\end{align}
where $1\le k,l \le3$.} 

\textcolor{blue}{
Then we derive $\bar{\mathbf{J}}_{\tilde{\mathbf{w}}}$. Define ${\mathbf{c}}_{\bar{\mathbf{\Theta }}\left( n,t \right)}=[ {c}_{\alpha \left( n,t \right)}, 0, {\mathbf{c}}^\top_{\boldsymbol{\eta}\left( n,t \right)} ] ^{\top}$, where ${c}_{\alpha \left( n,t \right)}$ is given in (\refeq{c_alpha}). Combining (\refeq{DdRU_eta_near}), the $k$-th element of ${\mathbf{c}}_{\boldsymbol{\eta}\left( n,t \right)}$ is defined as
\begin{align}
    \frac{\partial \mathbf{C}_{\tilde{\mathbf{w}}_{n,t}}}{\partial \left[ \boldsymbol{\eta } \right] _k}=2\alpha ^2\frac{\sqrt{2P_{\mathrm{t}}}\lambda _n}{4\pi}\left| x_{n,t} \right|^2\boldsymbol{\gamma }_{\mathrm{RU},n}^{\mathrm{T}}\left( \dot{\mathbf{d}}_{\mathrm{RU}}\circledast \frac{\partial \mathbf{d}_{\mathrm{RU}}}{\partial \left[ \boldsymbol{\eta } \right] _k} \right) \sigma _{\mathrm{H}}^{2}\mathbf{I}\triangleq \left[ {\mathbf{c}}_{\boldsymbol{\eta }\left( n,t \right)} \right] _k\sigma _{\mathrm{H}}^{2}\mathbf{I}.
\end{align}
Therefore, $\bar{\mathbf{J}}_{\tilde{\mathbf{w}}}$ is given by
\begin{align}
    \bar{\mathbf{J}}_{\tilde{\mathbf{w}}}=\sum_{t=1}^T{\sum_{n=1}^N{\frac{N_{\mathrm{B}}\sigma _{\mathrm{H}}^{4}}{\left( \alpha ^2|x_{n,t}|^2\left\| \boldsymbol{\gamma }_{\mathrm{RU},n} \right\| ^2\sigma _{\mathrm{H}}^{2}+\sigma ^2 \right) ^2}}}{\mathbf{c}}_{\bar{\mathbf{\Theta}}\left( n,t \right)}{\mathbf{c}}_{\bar{\mathbf{\Theta}}\left( n,t \right)}^{\top}.
\end{align}
}
Unlike the traditional analysis that the CRLBs for the intermediate parameters $d_\mathrm{RU}$, $\varphi_\mathrm{RU}$ and $\theta_\mathrm{RU}$ are obtained by extracting the corresponding diagonal element of \textcolor{blue}{$\bar{\boldsymbol{\mathcal{J}}}^{-1}$}, we adopt the notion of EFI to evaluate the estimation performance limits of the intermediate parameters \cite{EFIM}. First define 
\begin{align}
    \bar{E}(k)\triangleq [\textcolor{blue}{\bar{\boldsymbol{\mathcal{J}}}}]_{k,k}-[\textcolor{blue}{\bar{\boldsymbol{\mathcal{J}}}}]_{k,\left( 1:5 \right) \setminus k}[\textcolor{blue}{\bar{\boldsymbol{\mathcal{J}}}}]_{\left( 1:5 \right) \setminus k,\left( 1:5 \right) \setminus k}^{-1}[\textcolor{blue}{\bar{\boldsymbol{\mathcal{J}}}}]_{\left( 1:5 \right) \setminus k,k},
\end{align}
where $k\in\{3,4,5\}$. The index set $(1:5)\backslash k$ contains all the indexes from $1$ to $5$ except index $k$. Then the EFI for $d_\mathrm{RU}$, $\varphi_\mathrm{RU}$ and $\theta_\mathrm{RU}$
% , and the relationship between the EFI and $\bar{\mathbf{J}}^{-1}$  
are  given by \cite{EFIM}
% \begin{subequations}
%     \label{EFI}
%     \begin{align}
%         \bar{J}_\mathrm{E}(d_\mathrm{RU})=\bar{E}(3)=\frac{1}{[\bar{\mathbf{J}}^{-1}]_{3,3}}; \\
%         \mspace{1mu} \bar{J}_\mathrm{E}(\varphi_\mathrm{RU})=\bar{E}(4)=\frac{1}{[\bar{\mathbf{J}}^{-1}]_{4,4}};\\
%         \mspace{1mu} \bar{J}_\mathrm{E}(\theta_\mathrm{RU})=\bar{E}(5)=\frac{1}{[\bar{\mathbf{J}}^{-1}]_{5,5}}. 
%     \end{align}
% \end{subequations}
    \begin{align}
        \label{EFI}
        \bar{J}_\mathrm{E}(d_\mathrm{RU})=\bar{E}(3)=\frac{1}{[{\textcolor{blue}{\bar{\boldsymbol{\mathcal{J}}}}}^{-1}]_{3,3}}; \quad \bar{J}_\mathrm{E}(\varphi_\mathrm{RU})=\bar{E}(4)=\frac{1}{[{\textcolor{blue}{\bar{\boldsymbol{\mathcal{J}}}}}^{-1}]_{4,4}};\quad \bar{J}_\mathrm{E}(\theta_\mathrm{RU})=\bar{E}(5)=\frac{1}{[{\textcolor{blue}{\bar{\boldsymbol{\mathcal{J}}}}}^{-1}]_{5,5}}. 
    \end{align}
Note that the EFI given in (\refeq{EFI}) quantifies the information limits about the intermediate parameters that could be attained from the received samples.

\section{The Impact of Each Part of the Cascaded Channel on the Fundamental Limits}\label{sec:IV}
In this section, we analyze how the three parts of the cascaded channel, namely  $\mathbf{h}_{\mathrm{RU},n}$, $\mathbf{H}_{\mathrm{BR},n}$ and $\mathbf{\Phi}_t$, influence the fundamental limits of the near-field localization. Based on the analysis and the results in Section \ref{sec:results}, we investigate the near-field effect in both RIS-UE and BS-RIS part of the channel, and compare the results with those of the far-field scenario. Besides, we also reveal that the well-known focusing control scheme for the RIS may degrade the localization performance in the asynchronous scenario.

% we first prove that when near-field asynchronous scenario is considered for the RIS-UE part of the channel, it is theoretically possible to localize the UE since the information for both the distance and the direction can be obtained, while in the far-field asynchronous scenario, the distance parameter is impossible to estimate and only the direction of the UE can be inferred. Besides, we reveal that whether the BS-RIS part of the channel works in the near-field or far-field scenario will cause different influence on the PEB performance. The BS antenna array will provide different types of the performance gain in the two scenarios. Finally, we will show that it is not always a good choice to take maximizing the SNR as the RIS phase design criterion since it may degrade the localization performance in the asynchronous scenario.  
\subsection{Near-field vs. Far-field: RIS-UE Part}\label{sec:RIS-UE}
First we focus on the channel between the UE and the RIS. In this part of the channel, when the spherical wavefront is considered as (\refeq{dru_near}), (\refeq{DdRU_pU_near}) and (\refeq{DdRU_eta_near}), both the FIM \textcolor{blue}{${\boldsymbol{\mathcal{J}}}$ and $\bar{\boldsymbol{\mathcal{J}}}$} are generally invertible and the EFI for the intermediate parameters (\refeq{EFI}) are positive with adequate samples, which means that theoretically the position of the UE and all the intermediate parameters related to the position can be estimated in the near-field scenario. \textcolor{blue}{That is, although the TOA information cannot be inferred due to the synchronization mismatch $\xi$ and the unknown attenuation $\alpha$, the position information of the UE can be obtained from the curvature of the spherical wavefront \emph{passively} sensed by the RIS.\footnote{\textcolor{blue}{Unlike the traditional near-field localization scenario that the COA is sensed by the \emph{active} antenna array.}}} However, as the distance between the user and the RIS $d_\mathrm{RU}\to\infty$ or when the far-field planar wavefront is adopted as (\refeq{drU_far}), we will prove that it is impossible to localize the UE because the position-related parameter $d_\mathrm{RU}$ cannot be estimated.
\begin{proposition}
    \label{prop.1}
    When $d_\mathrm{RU}\to\infty$ or the far-field scenario is considered in the RIS-UE part of the channel, the EFI for the distance parameter $\bar{J}_\mathrm{E}(d_\mathrm{RU})=0$, \textcolor{blue}{regardless of whether considering the multi-paths between the BS and the RIS.}
\end{proposition}
\begin{IEEEproof}[\bf{Proof}]
    \color{blue}
  Please refer to Appendix \ref{appendix1}.
\end{IEEEproof}

When $d_\mathrm{RU}\to\infty$, the wavefront tends to be planar as the far-field scenario. According to Proposition \ref{prop.1}, in this case we are unable to obtain any information about the distance parameter $d_\mathrm{RU}$ with the existence of the synchronization mismatch $\xi$. Because in the near-field scenario, the values of $\frac{\partial {d}_{r\mathrm{U}}}{\partial d_{\mathrm{RU}}}$ are different across different $r$, the RIS is able to collect the information about $d_\mathrm{RU}$ from the spherical wavefront through the phase differences \textcolor{blue}{and the amplitude differences} caused by $d_\mathrm{RU}$ at different RIS elements. However, when the wavefront tends to be planar, from (\refeq{DdrU_dRU_far}) $\frac{\partial {d}_{r\mathrm{U}}}{\partial d_{\mathrm{RU}}}\to 1$ for all $r$, which means that in the far-field scenario the RIS cannot collect the information about the parameter $d_\mathrm{RU}$ since no phase difference \textcolor{blue}{or amplitude difference} caused by $d_\mathrm{RU}$ exists. 

It is worth underlining that the direction of  the UE can still be estimated when the wavefront tends to be planar. 
In the far field scenario,  $\mu_{b,n,t}$ can be expressed as
\begin{align}
    \color{blue}
    \mu _{b,n,t}\simeq \alpha x_{n,t}e^{-j2\pi f_n(c\xi +d_{\mathrm{RU}})/c}\gamma _{\mathrm{RU},n}\boldsymbol{\phi }_{t}^{\top}(\mathbf{h}_{b\mathrm{R},n}\circledast \hat{\mathbf{h}}_{\mathrm{RU},n}),
\end{align}
where $[\hat{\mathbf{h}}_{\mathrm{RU},n}]_r = e^{-j2\pi f_n\Gamma_{r\mathrm{U}}/c}$ for $1\le r\le N_\mathrm{R}$.
It can be drawn (as in Section \ref{sec:numerical results}) that, when investigating the following parameter $\bar{\mathbf{\Theta}}'=\left[ \alpha , c\xi +d_{\mathrm{RU}}, \varphi _{\mathrm{RU}}, \theta _{\mathrm{RU}} \right] ^{\top} \in \mathbb{R}^{4\times1}$, the EFI for the direction related parameters $\varphi_\mathrm{RU}$ and $\theta_\mathrm{RU}$
are generally positive in \emph{both} near-field and far-field scenarios. When $d_\mathrm{RU}\to \infty$ or the planar wavefront is considered, $\frac{\partial {d}_{r\mathrm{U}}}{\partial \varphi _{\mathrm{RU}}}$ and $\frac{\partial {d}_{r\mathrm{U}}}{\partial \theta _{\mathrm{RU}}}$ are still different for different $r$ as:
\begin{align}
    % \frac{\partial {d}_{r\mathrm{U}}}{\partial \varphi _{\mathrm{RU}}} &\to x_{\mathrm{R},r}\sin \theta _{\mathrm{RU}}\sin \varphi _{\mathrm{RU}}-y_{\mathrm{R},r}\sin \theta _{\mathrm{RU}}\cos \varphi _{\mathrm{RU}}  , \nonumber\\
    % \frac{\partial {d}_{r\mathrm{U}}}{\partial \theta _{\mathrm{RU}}}&\to -x_{\mathrm{R},r}\cos \theta _{\mathrm{RU}}\cos \varphi _{\mathrm{RU}}-y_{\mathrm{R},r}\cos \theta _{\mathrm{RU}}\sin \varphi _{\mathrm{RU}} \nonumber \\
    % &\pheq\;+z_{\mathrm{R},r}\sin \theta _{\mathrm{RU}}.
    \frac{\partial {d}_{r\mathrm{U}}}{\partial \varphi _{\mathrm{RU}}} &\to x_{\mathrm{R},r}\sin \theta _{\mathrm{RU}}\sin \varphi _{\mathrm{RU}}-y_{\mathrm{R},r}\sin \theta _{\mathrm{RU}}\cos \varphi _{\mathrm{RU}}  , \nonumber\\
    \frac{\partial {d}_{r\mathrm{U}}}{\partial \theta _{\mathrm{RU}}}&\to -x_{\mathrm{R},r}\cos \theta _{\mathrm{RU}}\cos \varphi _{\mathrm{RU}}-y_{\mathrm{R},r}\cos \theta _{\mathrm{RU}}\sin \varphi _{\mathrm{RU}}+z_{\mathrm{R},r}\sin \theta _{\mathrm{RU}}.
\end{align}
Therefore, unlike the distance parameter $d_\mathrm{RU}$, the phase differences caused by $\varphi _{\mathrm{RU}}$ and $\theta _{\mathrm{RU}}$ exist at different RIS elements in both near-field and far-field scenarios. 

\subsection{Near-field vs. Far-field: BS-RIS Part}\label{sec:BS-RIS}
\subsubsection{Spatial Gain vs. Power Gain}
Unlike the traditional localization scenario that the BS, equipped with an active large-scale antenna array, directly receives the signal transmitted from the UE, there is an extra part of the channel $\mathbf{H}_{\mathrm{BR},n}$ between the BS and the RIS cascaded with $\mathbf{h}_{\mathrm{RU},n}$ in the RIS-assisted localization scenario. In this subsection, we investigate the impact of the BS-RIS part of the channel on the localization CRLB in both near-field and far-field scenarios. This part of the channel is assumed to be determined since both the BS position and the RIS position are known in advance. The information of the UE position is not directly contained in this part of the channel. However, in the following discussions, we will reveal that this part of the channel determines the type of the gain provided by the BS antenna array. 

Let $\mathbf{J}^{(\mathrm{B})}_{\boldsymbol{\mu}b} = \sum_{t=1}^T{\sum_{n=1}^N}\mathbf{J}_{b,n,t}$ denote the Fisher information provided by all time slots and sub-carriers at antenna $b$, then we have \textcolor{blue}{$\mathbf{J}_{\boldsymbol{\mu}} = \sum_{b=1}^{N_\mathrm{B}} \mathbf{J}^{(\mathrm{B})}_{\boldsymbol{\mu}b}$}. According to the discussion in Remark \ref{remark1}, we know that $\mathbf{J}_{\boldsymbol{\mu}} \succeq \mathbf{J}^{(\mathrm{B})}_{\boldsymbol{\mu}b}$, thus\footnote{When $\mathbf{J}^{(\mathrm{B})}_{\boldsymbol{\mu}b}$ and $\mathbf{J}_{\boldsymbol{\mu}} $ are invertible.} $\mathrm{PEB}(\mathbf{J}_{\boldsymbol{\mu}} )\le  \mathrm{PEB}(\mathbf{J}^{(\mathrm{B})}_{\boldsymbol{\mu}b})$  for any $b$, which indicates that the multiple antennas at the BS can improve the CRLB performance. In the near-field scenario as (\refeq{HBR_near}), $\mathbf{J}^{(\mathrm{B})}_{\boldsymbol{\mu}b}$ is generally different for different $b$, that is, each antenna at the BS can provide independent \emph{spatial gain} for the FIM. While in the far-field scenario, we will show in Proposition \ref{prop.2} that the multiple antennas at the BS only provide the \emph{power gain} for the FIM. That is, in this case the PEB decreases linearly with $\sqrt{N_\mathrm{B}}$, \textcolor{blue}{regardless of whether considering the multi-paths between the BS and the RIS.}  
\begin{proposition}\label{prop.2}
    When the far-field scenario is considered as (\refeq{HBR_far}) for \textcolor{blue}{$\mathbf{H}_{\mathrm{BR},n}^{\mathrm{(d)}}$, $\mathbf{J}^{(\mathrm{B})}_{\boldsymbol{\mu}b}$} is identical for all $b$, and 
    \textcolor{blue}{$\mathrm{PEB}(\boldsymbol{\mathcal{J}} )\propto \frac{1}{\sqrt{N_{\mathrm{B}}}}$}.
\end{proposition}
\begin{IEEEproof}[\bf{Proof}]
    \color{blue}
   Please refer to Appendix \ref{appendix2}.
\end{IEEEproof}

\subsubsection{Case Investigation}
To vividly show the different impacts of the two types of the gain on the CRLB performance, we then investigate the following case. Suppose that only a single carrier with index $n_0$ is used and the signal lasts only a single time slot $t_0$, during which the reflection coefficients of the RIS are unchanged. In this case $\mathbf{J}^{(\mathrm{B})}_{\boldsymbol{\mu}b}=\mathbf{J}_{b,n_0,t_0}$, which makes $\mathbf{J}^{(\mathrm{B})}_{\boldsymbol{\mu}b}$ a singular matrix with rank $2$ \cite{2022WangZY}. \textcolor{blue}{Besides, in this case
\begin{align}
    \mathbf{J}_{\tilde{\mathbf{w}}}=\frac{\sigma _{\mathrm{H}}^{4}}{\left( \alpha ^2|x_{n_0,t_0}|^2\left\| \boldsymbol{\gamma }_{\mathrm{RU},n} \right\| ^2\sigma _{\mathrm{H}}^{2}+\sigma ^2 \right) ^2}{\mathbf{c}}_{\mathbf{\Theta }\left( n_0,t_0 \right)}{\mathbf{c}}_{\mathbf{\Theta }\left( n_0,t_0 \right)}^{\top}
\end{align}
which has rank 1. Thus $\boldsymbol{\mathcal{J}}$ is also singular, which means that it is impossible to localize the UE through a single antenna under these circumstances.} Moreover, when the far-field scenario is considered for \textcolor{blue}{$\mathbf{H}_\mathrm{BR,n_0}^{(\mathrm{d})}$,  $\mathbf{J}_{\boldsymbol{\mu}}=N_\mathrm{B}\mathbf{J}^{(\mathrm{B})}_{\boldsymbol{\mu}b}$ is also a rank 2 singular matrix.} In other words, even if we collect the information from all antennas, we are still unable to localize the UE because in this scenario the gain provided by the antenna array at the BS for the localization is power gain, which can not increase the rank of the FIM. However, in the near-field scenario for \textcolor{blue}{$\mathbf{H}_\mathrm{BR,n_0}^{(\mathrm{d})}$}, the FIM \textcolor{blue}{$\mathbf{J}_{\boldsymbol{\mu}} = \sum_{b=1}^{N_\mathrm{B}} \mathbf{J}^{(\mathrm{B})}_{\boldsymbol{\mu}b}$} will generally become a full rank invertible matrix with adequate antennas.
In this scenario, the gain provided by the antenna array at the BS for the localization is defined as spatial gain, which can increase the rank of the FIM.  Therefore, unlike the far-field scenario, in this case we are theoretically possible to localize the UE. 

We can understand the above results from another perspective of view. Note that \textcolor{blue}{$\mathbf{H}_{\mathrm{BR},n_0}^{(\mathrm{d})}$} is irrelevant to $\mathbf{\Theta}$, then the Jacobian matrix $ \frac{\partial \boldsymbol{\mu }_{n_0,t_0}}{\partial \mathbf{\Theta }^{\top}}\in\mathbb{C}^{N_\mathrm{B}\times 5}$  can be expressed as
\begin{align}
    \color{blue}
    \frac{\partial \boldsymbol{\mu }_{n_0,t_0}}{\partial \mathbf{\Theta }^{\top}}=\mathbf{H}_{\mathrm{BR},n_0}^{(\mathrm{d})}\frac{\partial \alpha x_{n,t}e^{-j2\pi f_n\xi}\mathbf{\Phi }_{t_0}\mathbf{\Upsilon }_{\mathrm{RU},n_0}\mathring{\mathbf{h}}_{\mathrm{RU},n_0}}{\partial \mathbf{\Theta }^{\top}}.
\end{align}
In this case, \textcolor{blue}{$\mathbf{J}_{\boldsymbol{\mu}}$} can be rewritten as 
\begin{align}
    \label{FIM_case}
    \mathbf{J}_{\boldsymbol{\mu}}=\frac{2}{\sigma^2}\Re \left\{  \frac{\partial \boldsymbol{\mu }_{n_0,t_0} ^{\mathrm{H}}}{\partial \mathbf{\Theta }} \frac{\partial \boldsymbol{\mu }_{n_0,t_0}}{\partial \mathbf{\Theta }^{\top}} \right\}.
\end{align}
In the far-field scenario, obviously $\frac{\partial \boldsymbol{\mu }_{n_0,t_0}}{\partial \mathbf{\Theta }^{\top}}$ has rank $1$ since \textcolor{blue}{$\mathbf{H}_{\mathrm{BR},n_0}^{(\mathrm{d})}$} has rank $1$, according to (\refeq{HBR_far}). Therefore, from (\refeq{FIM_case}), again we find that \textcolor{blue}{$\mathbf{J}_{\boldsymbol{\mu}}$} is a singular matrix with rank $2$,  while in the near-field scenario, \textcolor{blue}{$\mathbf{H}_{\mathrm{BR},n_0}^{(\mathrm{d})}$} has full rank $N_\mathrm{B}$ (assuming $N_\mathrm{B}<N_\mathrm{R}$). Therefore, with large enough $N_\mathrm{B}$, $\frac{\partial \boldsymbol{\mu }_{n_0,t_0}}{\partial \mathbf{\Theta }^{\top}}$ will also generally become a full rank matrix with rank $5$, which eventually makes \textcolor{blue}{$\mathbf{J}_{\boldsymbol{\mu}}$ and the FIM $\boldsymbol{\mathcal{J}}$ invertible.} From this case, we find that the rank of the BS-RIS part of the channel matrix \textcolor{blue}{$\mathbf{H}_{\mathrm{BR},n_0}^{(\mathrm{d})}$} plays an important role.
\textcolor{blue}{
\begin{proposition}
    In the single-carrier, single snapshot case, it requires \emph{both} the BS-RIS part and the RIS-UE part of the channel work in the near-field scenario to localize the UE. The near-field effect in the RIS-UE part of the channel makes the RIS could passively sense the UE position through COA information, while the near-field effect in the BS-RIS part of the channel makes the multiple BS antennas provide sufficient degrees of freedom to extract the COA information.
\end{proposition}}
\begin{remark}\label{remark2}
    In this case where the information received by a single antenna is insufficient to estimate  $\boldsymbol{\Theta}$, the power gain in the far-field scenario does not improve the CRLB performance, while the spatial gain does. Indeed, for the same system parameters, when there is no much information about $\mathbf{\Theta}$ (provided by all the time slots and sub-carriers) at each single antenna, the spatial gain usually achieves a better CRLB performance than the power gain. 
    % the BS antenna array which could provide the spatial gain usually leads to better CRLB performance compared with the one only provides the power gain. 
    However, when adequate information about $\mathbf{\Theta}$ can be obtained from a single BS antenna, the power gain may outperform the spatial gain, which will be shown in Section \ref{sec:numerical results}. Besides, in the practical situation when the positions of the BS and the RIS are fixed, we can determine the gain type we prefer  achieving by adjusting the antenna separation at the BS. We can reduce the antenna separation if we prefer achieving the power gain, while we can increase the antenna separation if the effect of the spatial gain is preferred.
    % We tend to achieve the power gain with small antenna separation, while we tend to achieve the spatial gain when the antenna separation is increased.
\end{remark}

\subsection{The Impact of the RIS Coefficients}\label{sec:IV-C}
The reflection coefficient profile of the RIS, namely $\boldsymbol{\phi}_t$, is another important parameter that can significantly influence the CRLB performance. 
The random phase profile has been adopted and evaluated in many existing works \cite{2022WangZY}–\cite{2021Abu-Shaban}  since it is simple to implement. However, as pointed out in these works, the random phase profile usually cannot perform as well as the dedicated phase profile which is designed according to certain criterion. \textcolor{blue}{The focusing control plan, which takes maximizing the received SNR as the design criterion, has been investigated not only in the traditional wireless communications as a pre-coding scheme \cite{2022Zhang} but also in the RIS-assisted localization \cite{2021Elzanaty}–\cite{2021Abu-Shaban} as a RIS control scheme.} This efficient scheme avoids tackling the complex high-dimensional non-convex problem of directly optimizing the RIS configuration for the minimum PEB. The optimal SNR will lead to optimal spectral efficiency, which usually means better communication performance \cite{2022WangZY}. In the localization scenario investigated in this paper, however, the optimal SNR (or in other words the maximum RIS gain) does not ensure the optimal CRLB performance. In fact, in some cases, \textcolor{blue}{we could barely obtain any information about the UE's position can be obtained when the maximum received SNR is achieved with focusing scheme.} 

Denote $\bar{{\mathbf{J}}}_\mathrm{E}(\boldsymbol{\eta})$ as the EFIM for the position related intermediate parameters $\boldsymbol{\eta}$, given by
\begin{align}
    \label{JE_eta}
    \color{blue}
    \bar{{\mathbf{J}}}_\mathrm{E}(\boldsymbol{\eta})=[\bar{\boldsymbol{\mathcal{J}}}]_{3:5,3:5}-[\bar{\boldsymbol{\mathcal{J}}}]_{3:5,1:2}[\bar{\boldsymbol{\mathcal{J}}}]_{1:2,1:2}^{-1}[\bar{\boldsymbol{\mathcal{J}}}]_{1:2,3:5}.
\end{align} 
\textcolor{blue}{
Let $\mathbf{h}_{b\mathrm{R},n}=\boldsymbol{\gamma }_{b\mathrm{R},n}\circledast \mathring{\mathbf{h}}_{b\mathrm{R},n}$, where $[\boldsymbol{\gamma }_{b\mathrm{R},n}]_r=\frac{\lambda _n}{4\pi d_{br}}$, $[\mathring{\mathbf{h}}_{b\mathrm{R},n}]_r=e^{-j2\pi f_nd_{br}/c}$ ($1\le r\le N_\mathrm{R}$).}
\begin{figure}
    \centering
    \hspace{-0.5cm}
	\subfloat[]{
		\includegraphics[width=4.5cm]{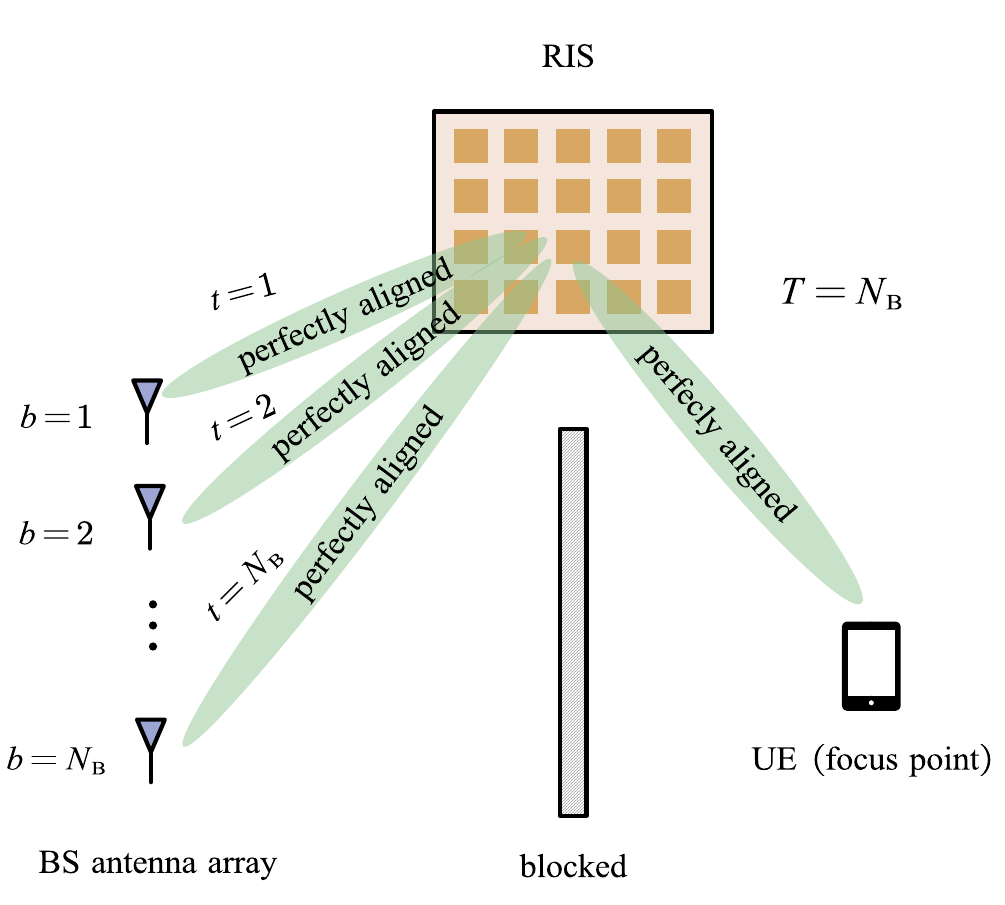}
	}
	\hspace{-0.4cm}
	\subfloat[]{
		\includegraphics[width=4.5cm]{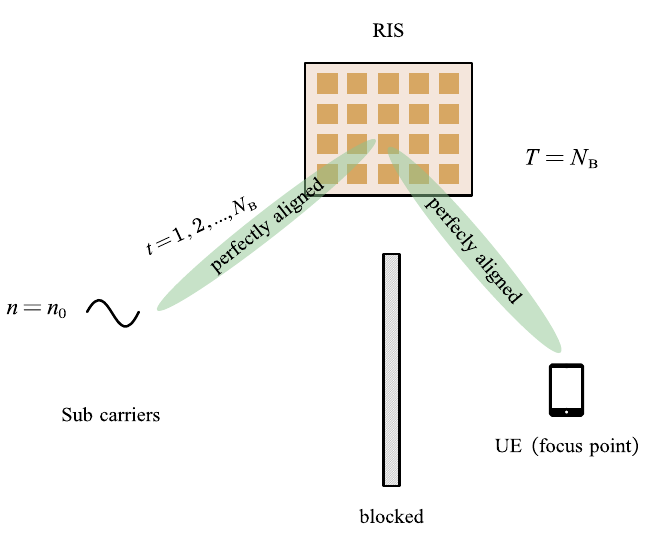}
	} 
    \hspace{-0.5cm}
    \subfloat[]{
		\includegraphics[width=4.5cm]{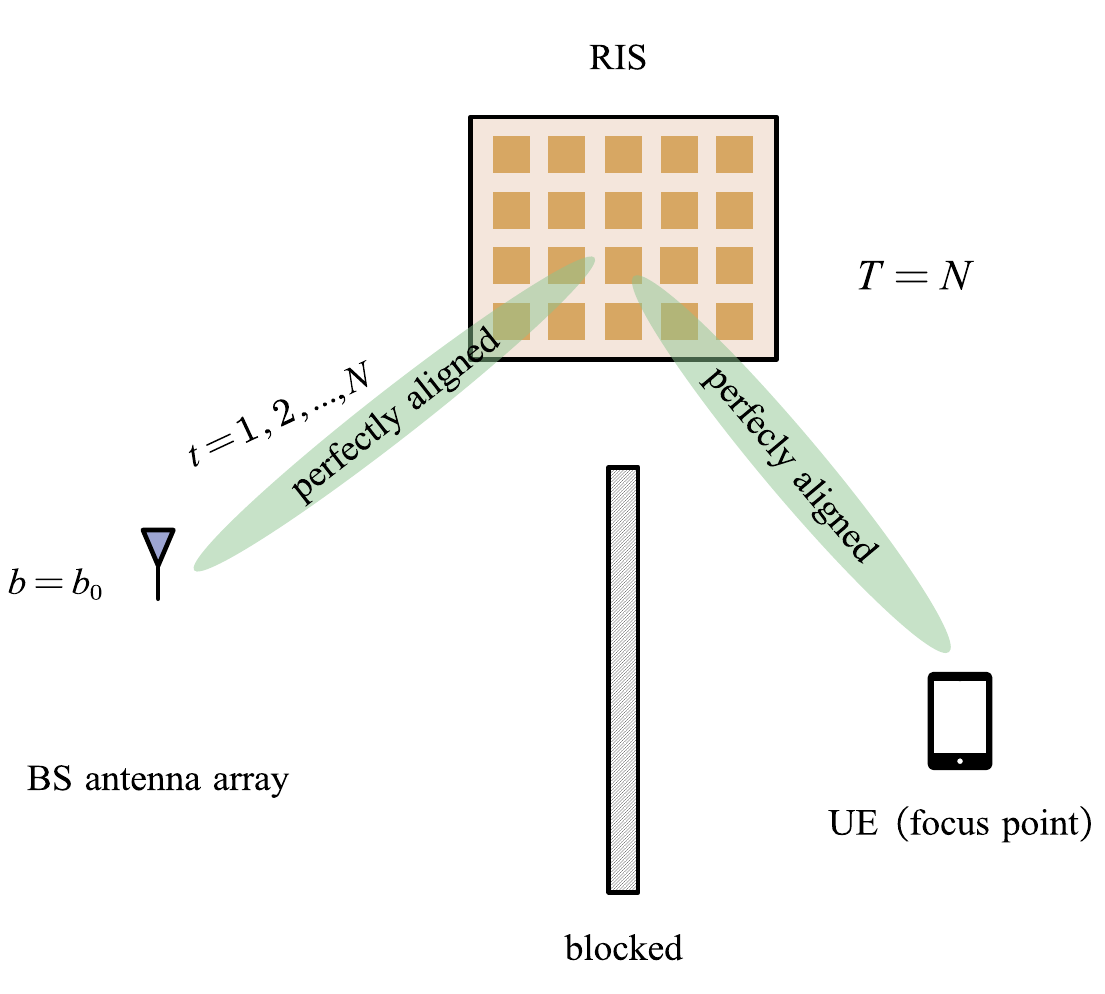}
	}
	\hspace{-0.4cm}
    \\
	\subfloat[]{
		\includegraphics[width=4.5cm]{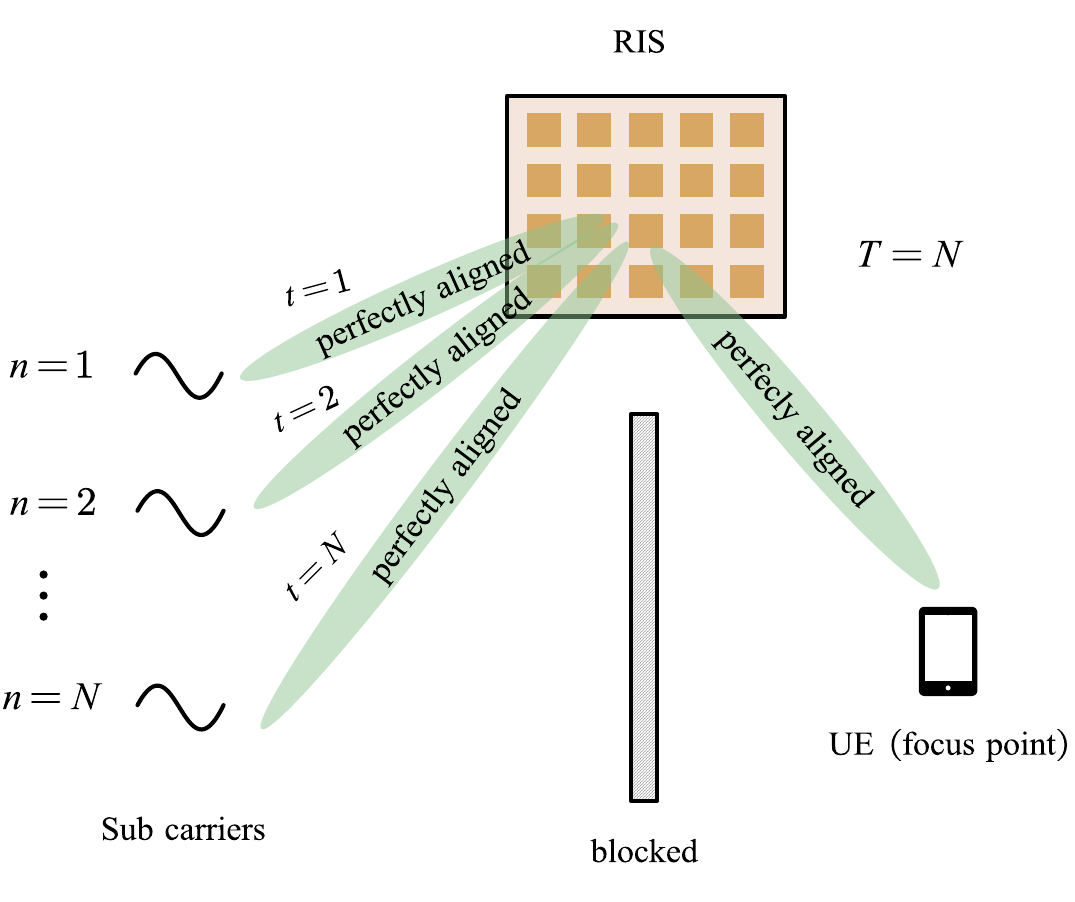}
	} 
    \hspace{-0.5cm}
    \subfloat[]{
		\includegraphics[width=4.5cm]{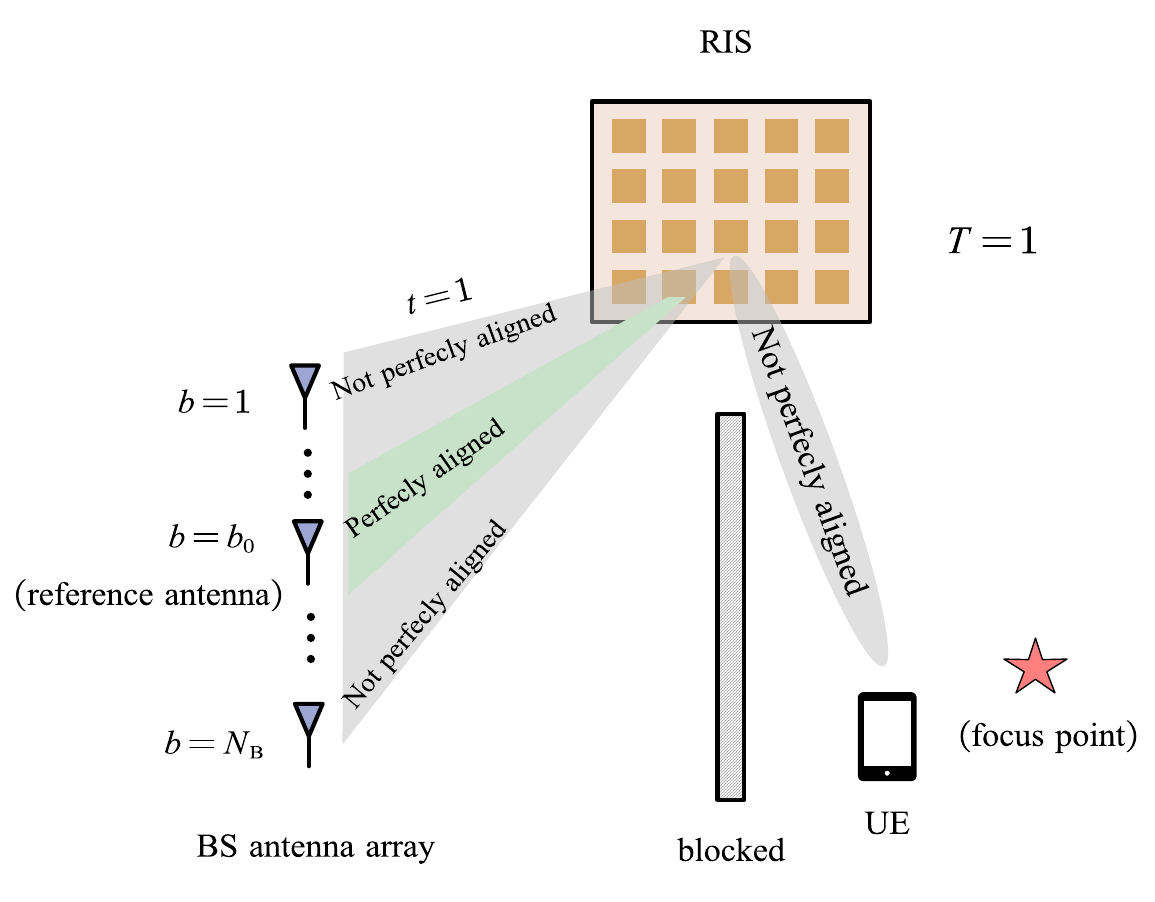}
	}
	\hspace{-0.3cm}
	\subfloat[]{
		\includegraphics[width=4.5cm]{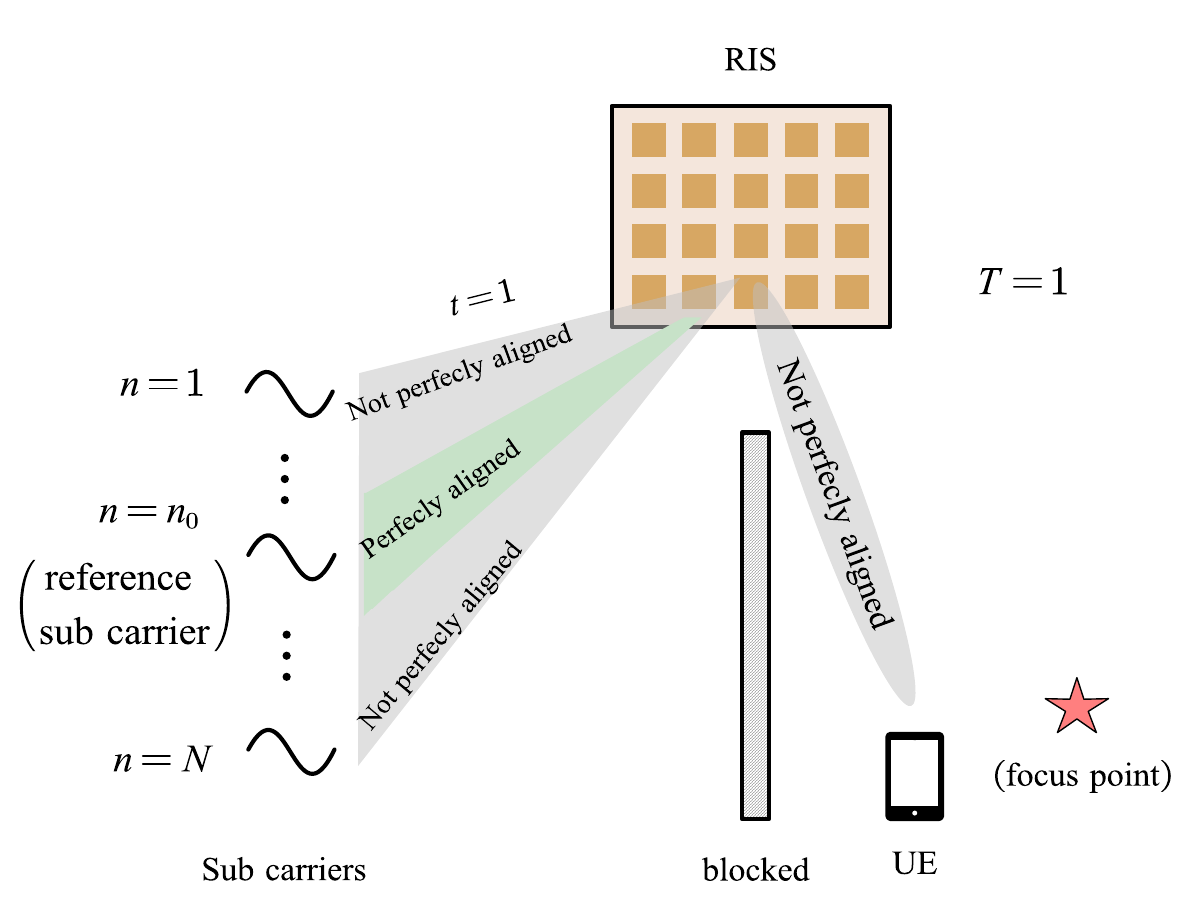}
	} 
    \hspace{-0.5cm}
    \caption{Schematic diagrams of Case 1, Case2, and the more practical situation. (a) Case1, SD. (b) Case1, FD. (c) Case2, SD. (d) Case2, FD. (e) Practical situation, SD. (f) Practical situation, FD. SD: Spatial domain. FD: Frequency domain.} \label{Fig.focuscases}
\end{figure}
\begin{proposition}\label{prop.4}
    \color{blue}
    When the multi-paths effect can be ignored between the BS and the RIS, in the following two cases $\mathbf{\bar{J}}_{\mathrm{E}}(\boldsymbol{\eta })\approx\mathbf{0}$.
    % \begin{itemize}
        % \item 

        \emph{Case 1:} Only a single carrier with index $n_0$ is used and $N_\mathrm{B}$ time slots are employed, and the size of the BS antenna array $L_\mathrm{B}\ll d_\mathrm{BR}$. The time varying phase profiles of the RIS are set as $\boldsymbol{\phi}_{t_1}=(\mathring{\mathbf{h}}_{t_1\mathrm{R},n_0}\circledast \mathring{\mathbf{h}}_{\mathrm{RU},n_0})^*$. 
        % \footnote{ Note that $\mathbf{\tilde{h}}^*_{b,n}$ is related to $\boldsymbol{\eta}$.}for $1\le t_1 \le N_\mathrm{B}$.
         Only the antenna with index $t_1$ is activated and the received signal is sampled during time slot $t_1$. 
        % \item 

        \emph{Case 2:} Only a single antenna with index $b_0$ is used and $N$ time slots are employed, and the bandwidth of the signal $f_N-f_c\ll f_c$. The time varying phase profiles of the RIS are set as $\boldsymbol{\phi}_{t_2}=(\mathring{\mathbf{h}}_{b_0\mathrm{R},t_2}\circledast \mathring{\mathbf{h}}_{\mathrm{RU},t_2})^*$ for $1\le t_2 \le N$. Only the received signal at sub-carrier $t_2$ is sampled during time slot $t_2$.
    % \end{itemize}
\end{proposition}
\begin{IEEEproof}[\bf{Proof}]
    \color{blue}
    Please refer to Appendix \ref{appendix3}.
\end{IEEEproof}

The schematic diagrams of Case1 and Case 2 in both spatial domain and frequency domain are shown in Figs. \ref{Fig.focuscases}(a)-(d). In Case 1, the RIS with coefficient $\boldsymbol{\phi}_{t_1}$ focuses the signal from the focus point $\mathbf{p}_\mathrm{U}$ to different antennas at different time slots. Therefore, all $N_\mathrm{B}$ antennas receive the signal with the maximum SNR in this case. The situation in Case 2 is similar, the  SNR of the received signal at each sub-carrier is maximized. However, the EFIM \textcolor{blue}{$\mathbf{\bar{J}}_{\mathrm{E}}(\boldsymbol{\eta })\approx\mathbf{0}$ in the two cases, which means that we can barely obtain any information about the direction and the distance of the UE, that is, the position of the UE cannot be effectively estimated.} The maximum received SNR, which usually means optimal communication performance, but leads to the worst localization performance in the two cases. The intuitive understanding of this result is that, \textcolor{blue}{when the unknown synchronization mismatch and the shadowing effect attenuation exist}, the position information of the UE is provided only by the spherical wavefront sensed by the RIS. Therefore, estimating the position related parameters (namely $d_{\mathrm{RU}}$, $\varphi _{\mathrm{RU}}$ and $\theta _{\mathrm{RU}}$) all relies on the \textcolor{blue}{phase differences and the amplitude differences at different RIS elements caused by these parameters.}
% Therefore, the phase differences at different RIS elements caused by the position related parameters (namely $d_{\mathrm{RU}}$, $\varphi _{\mathrm{RU}}$ and $\theta _{\mathrm{RU}}$) are significant in estimating these parameters. 
The focusing scheme, which achieves the maximum SNR (or equivalently the maximum RIS gain) by aligning the received signal at all RIS elements when observed either at each BS antenna in Case 1 or at each sub-carrier in Case 2, makes the phase differences vanish. \textcolor{blue}{The remaining amplitude differences information is not sufficient to effectively locate the UE because the amplitude differences across an array are far less significant compared with the phase differences}. Thus the position related parameters \textcolor{blue}{cannot be effectively estimated} in these circumstances.
\begin{remark}\label{remark3}
    In practical situations, we can get the signal from all antennas and sub-carries in a single time slot and usually the UE does not locate exactly at the focusing point (as shown in Figs. \ref{Fig.focuscases}(e)-(f)), which makes the EFIM  $\mathbf{\bar{J}}_{\mathrm{E}}(\boldsymbol{\eta })\succ\mathbf{0}$. However, what we want to reveal in Proposition \refeq{prop.4} is that the localization performance could be degraded with the increase of the received SNR, as will be verified in Section \ref{sec:numerical results}. In other words, maximizing the RIS gain, which ensures better communication performance, is not always a good objective for the asynchronous RIS-assisted localization. \textcolor{blue}{Besides, as will be shown in Section \ref{sec:numerical results}, this conclusion still holds when the multi-paths between the BS and the RIS cannot be ignored.}
    Further study is required on investigating how to efficiently set the RIS configuration to achieve an optimal localization performance in this scenario. \textcolor{blue}{Some insights are provided in Section \ref{sec:V-C}.}
\end{remark}

\section{Numerical Results}\label{sec:numerical results}
This section presents the numerical results to evaluate the performance limits of the RIS-assisted localization and the intermediate-parameter estimation. Besides, the properties revealed in Section \ref{sec:IV} are verified in this section. The 3D localization scenario is depicted in Fig. \ref{3Dscneario}. Unless indicated otherwise, some basic system parameters are set as follows. The carrier frequency is set as $f_c=28$ GHz. The reference points of the RIS and the BS are assumed to be located at $\mathbf{p}_\mathrm{R}=[0, 0, 0]^\top$(m) and $\mathbf{p}_\mathrm{B}=[8, -12, 2]^\top$(m), respectively. Both the RIS and the antenna array at the BS are lying on the Y-Z plane. The RIS is assumed to be a passive URA with $N_\mathrm{R}=60\times 60 =3600$ elements and the BS is equipped with an active URA with $N_\mathrm{B}=8\times 8 =64$ antennas. The spacing of two adjacent RIS elements is set as $\Delta r ={\lambda_c}/{2}$ where $\lambda_c$ is the carrier wavelength. \textcolor{blue}{The multi-paths between the BS and the RIS are considered in this section, the corresponding Rician factor $\kappa$ is set as $5\,$dB.}

% In order to conveniently show the results and the properties of the performance limits, we mainly investigate the performance limits of the UE locate at $z=-1$ plane in this section without loss of generality. 
% \begin{figure}[t]
%     \centering  
%     \hspace{-0.3cm}  
%     \includegraphics[width=9cm]{simu_PEB.pdf}
%     \caption{The PEB at $z=-1$ plane with random RIS phase profile. The received signal power is fixed at each point.}    \label{Fig.PEB} 
% \end{figure}
\begin{figure}[t] 
    \centering 
    \color{blue}
    \hspace{-0.3cm}   
    \includegraphics[width=8cm]{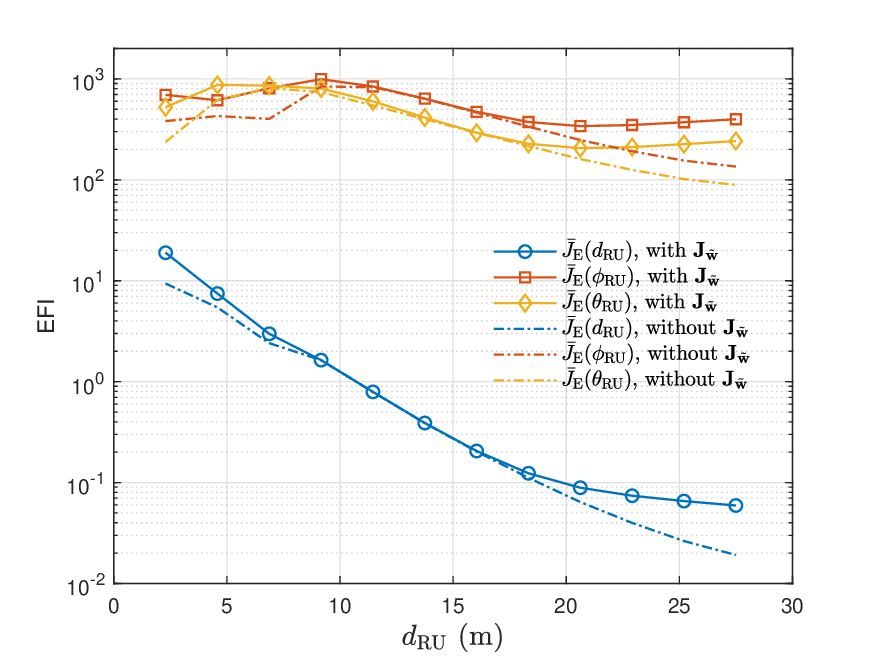}
    \caption{The EFI of intermediate parameters as a function of the distance between the RIS and the UE. The received signal power is fixed.}\label{Fig.EFI}
\end{figure}
\subsection{RIS-UE Part Near-Field Effect on PEB}
In this subsection, we focus on the near-field effect in the RIS-UE part of the channel. The power of the received signal from different locations is fixed in order to highlight the effect of the spherical wavefront. The random phase profile is adopted at the RIS, the bandwidth of the transmitted signal is set as $400$ MHz, and a single time slot is considered. 

% Fig. \refeq{Fig.PEB} depicts the PEB at $z=-1$ plane and 
Fig. \ref{Fig.EFI} shows the EFI of the position-related intermediate parameters as a function of $d_\mathrm{RU}$ \textcolor{blue}{with and without considering the multi-paths between the BS and the RIS.}
% in both near-field and far-field scenario. 
% Although the PEB pattern in Fig. \refeq{Fig.PEB} shows a certain degree of  randomness, it can still be observed that the positions with low PEB are mainly concentrated around the RIS (Region 1) and the positions far from the RIS are mostly with poor PEB performance (Region 2).
From Fig. \refeq{Fig.EFI}, with the increase of the distance between the RIS and the UE, the EFI of the distance parameter $d_\mathrm{RU}$ tends to $0$ \textcolor{blue}{regardless of whether considering the multi-paths between the BS and the RIS}, which is exactly the reason of the poor PEB when the UE is far from the RIS as discussed in Section \ref{sec:RIS-UE}. However, the EFI of the direction-related parameters $\theta_\mathrm{RU}$ and $\phi_\mathrm{RU}$ remains high regardless of $d_\mathrm{RU}$, which also verifies that the direction of arrival with high-accuracy is attainable in both scenarios. \textcolor{blue}{Besides, the EFI for all the position related parameters are generally improved when the multi-paths between the BS and the RIS are considered, which verifies the discussion in Remark \ref{remark1}.}

\begin{figure}[t] 
    \centering 
    \color{blue}
    \hspace{-1.9cm}
    \subfloat[]{
		\includegraphics[width=8cm]{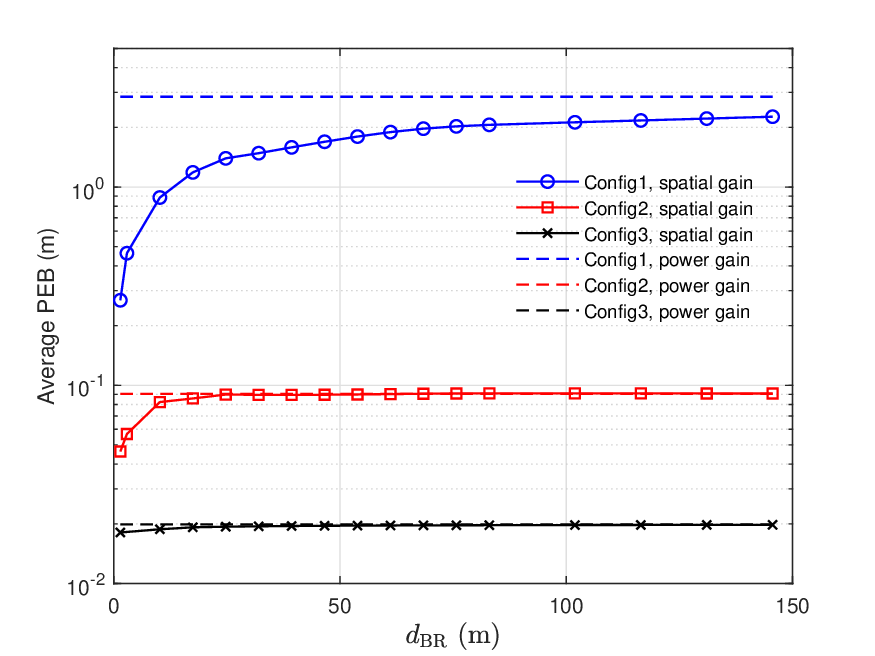}
	}
	\hspace{-0.8cm}\label{Fig.dBR}
    \subfloat[]{
		\includegraphics[width=8cm]{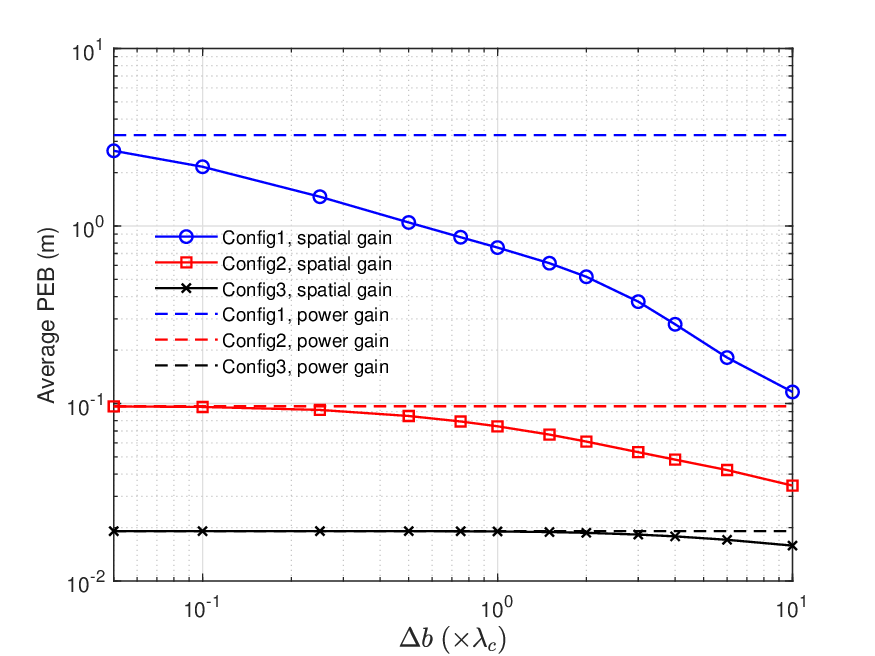}
	}
	\hspace{-1.9cm}\label{Fig.dB}
    \caption{Average PEB of 100 Mont Carlo trials with random RIS coefficients vs. $d_\mathrm{BR}$ and the BS antenna separation $\Delta b$. The UE is located in $\mathbf{p}_\mathrm{U}=[4, 2.1, -1]^\top$,  the received signal power is fixed at each point.}
\end{figure}
% \begin{figure}[!t] 
%     \centering 
%     \color{blue}
%     \hspace{-0.3cm}    
%     \includegraphics[width=9cm]{differentdB_average.eps}
%     \caption{Average PEB of 100 Mont Carlo trials with random RIS coefficients vs. the BS antenna separation $\Delta b$. The UE is located in $\mathbf{p}_\mathrm{U}=[4, 2.1, -1]^\top$, the received signal power is fixed at each point.}
% \end{figure}
\begin{figure}
    \centering
    \color{blue}
    \hspace{-0.5cm}
	\subfloat[]{
		\includegraphics[width=4cm]{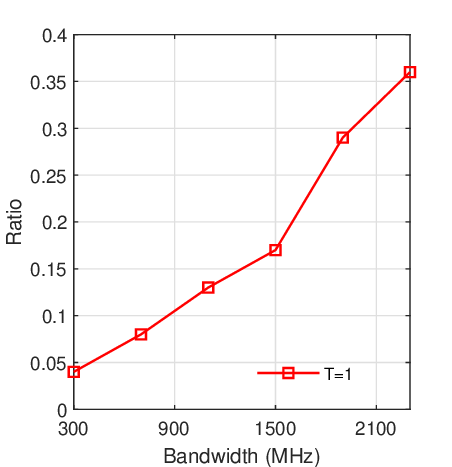}
	}
	\hspace{-0.5cm}\label{Fig.ratio1}
	\subfloat[]{
		\includegraphics[width=4cm]{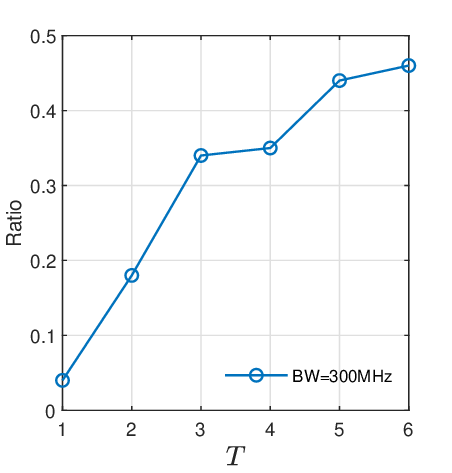}
	} 
    \hspace{-0.5cm}\label{Fig.ratio2}
    \caption{The ratio that the power gain outperforms the spatial gain within 100 Mont Carlo trials with random RIS coefficients. BW: bandwidth.} 
    \label{Fig.ratio}
\end{figure}
\subsection{Comparisons between the Spatial Gain and the Power Gain}
In this subsection, we focus on the BS-RIS part of the channel and compare the effect of the two different kinds of performance gains introduced in Section \ref{sec:BS-RIS} on PEB. 100 Mont Carlo trials with random RIS coefficients are carried out under different system configurations. 
\begin{table}[!ht]
	\renewcommand{\arraystretch}{1.3}
	\caption{A Notation of the System Configurations}
	\label{table1}
	\centering
    \begin{tabular}{ccc}
        \toprule
        Configuration & Bandwidth & Time slots \\
        \midrule
        Config1 & 300MHz & $T=1$ \\
        Config2 & 2500MHz & $T=1$\\
        Config3 & 2500MHz & $T=3$\\
        \bottomrule
    \end{tabular}
\end{table}

In Fig. \ref{Fig.dBR}, the received signal power is fixed and the spacing of two adjacent BS antennas is set as $\Delta b=\lambda_c/2$. The average PEBs of 100 Mont Carlo trials are depicted as a function of the distance between the BS and the RIS. Three system configurations are evaluated as listed in Table \ref{table1}. As expected, the spatial gain will approach the power gain as $d_\mathrm{BR}$ increases under all system configurations because the near-field model (\refeq{HBR_near}) will approach the far-field model (\refeq{HBR_far}) when $d_\mathrm{BR}\to \infty$. Besides, from the perspective of view of the average PEB, the spatial gain significantly outperforms the power gain when there is no much time and bandwidth resource at a single antenna (Config1). However, in Config2 and Config3, the performance gap tends to be negligible with the increase of bandwidth and time slots. Similar phenomena can be observed from Fig. \ref{Fig.dB}. The only difference is that, in Fig. \ref{Fig.dB}, without adjusting $d_\mathrm{BR}$, we could still let the spatial gain tend to approach the power gain by decreasing $\Delta b$, as discussed in Remark \ref{remark2}. In addition, from Fig. \ref{Fig.dB}, a small increase of the antenna separation $\Delta {b}$ could result in significant spatial gain when the time and bandwidth resources are deficient as Config1, whereas a tremendous increase of  $\Delta {b}$ is required to achieve observable spatial gain with abundant space and time resource as Config3.

To further demonstrate the impact of the bandwidth and time resources on the two different types of gains, Fig. \ref{Fig.ratio} depicts the ratio that the power gain outperforms the spatial gain within 100 Mont Carlo trials. From Fig. \ref{Fig.ratio}, the probability that the power gain is with better performance increases as the increase of bandwidth and time slots, which further verifies the discussion in Remark \ref{remark2}. This is because the spatial gain can be deemed as providing new degrees of freedom, while the power gain simply  `amplify' the information linearly. When the time and bandwidth resources are deficient, there is no much information about the position of the UE at a single antenna. Therefore, obtaining more degrees of freedom is certainly more effective than simply linearly `amplifying' such information. However, with the increase of time and bandwidth resources, the information of the UE position that we are able to collect from a single antenna becomes sufficient to accurately localize the UE. In this situation, `amplifying' such abundant information becomes profitable, while the influence of providing more degrees of freedom becomes limited.

\begin{figure*}[!t] 
    \centering 
    \color{blue}
    \hspace{-0.8cm}       
    \subfloat[]{
		\includegraphics[width=5.5cm]{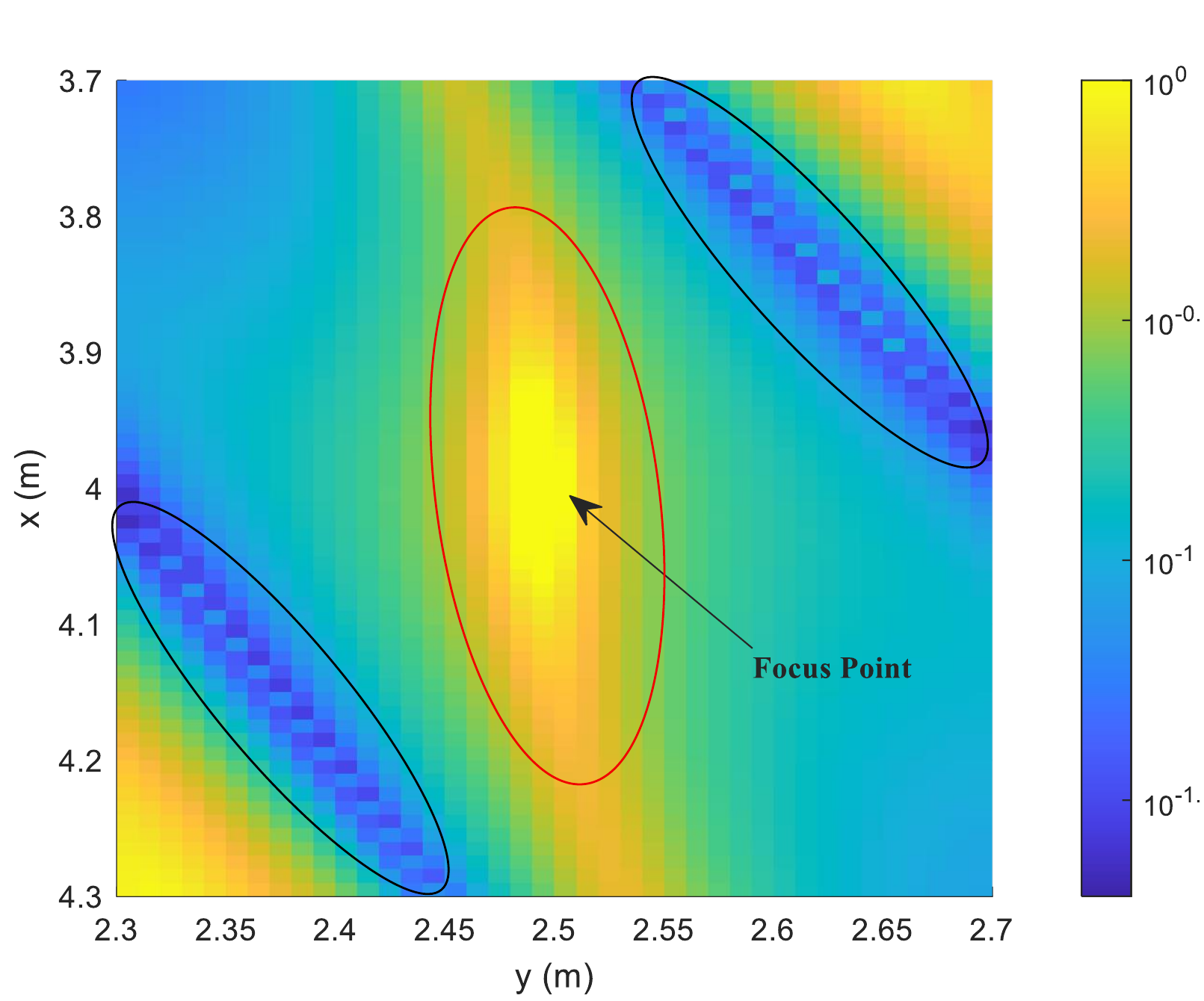}
	}    
	\hspace{-0.3cm}
    \subfloat[]{
		\includegraphics[width=5.5cm]{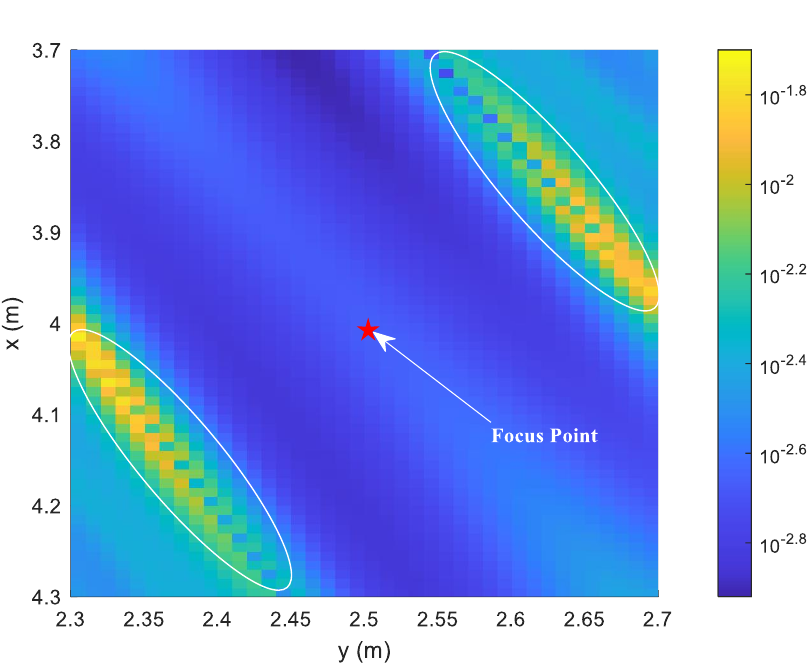}
	}
	\hspace{-0.3cm}
	\subfloat[]{
		\includegraphics[width=5.5cm]{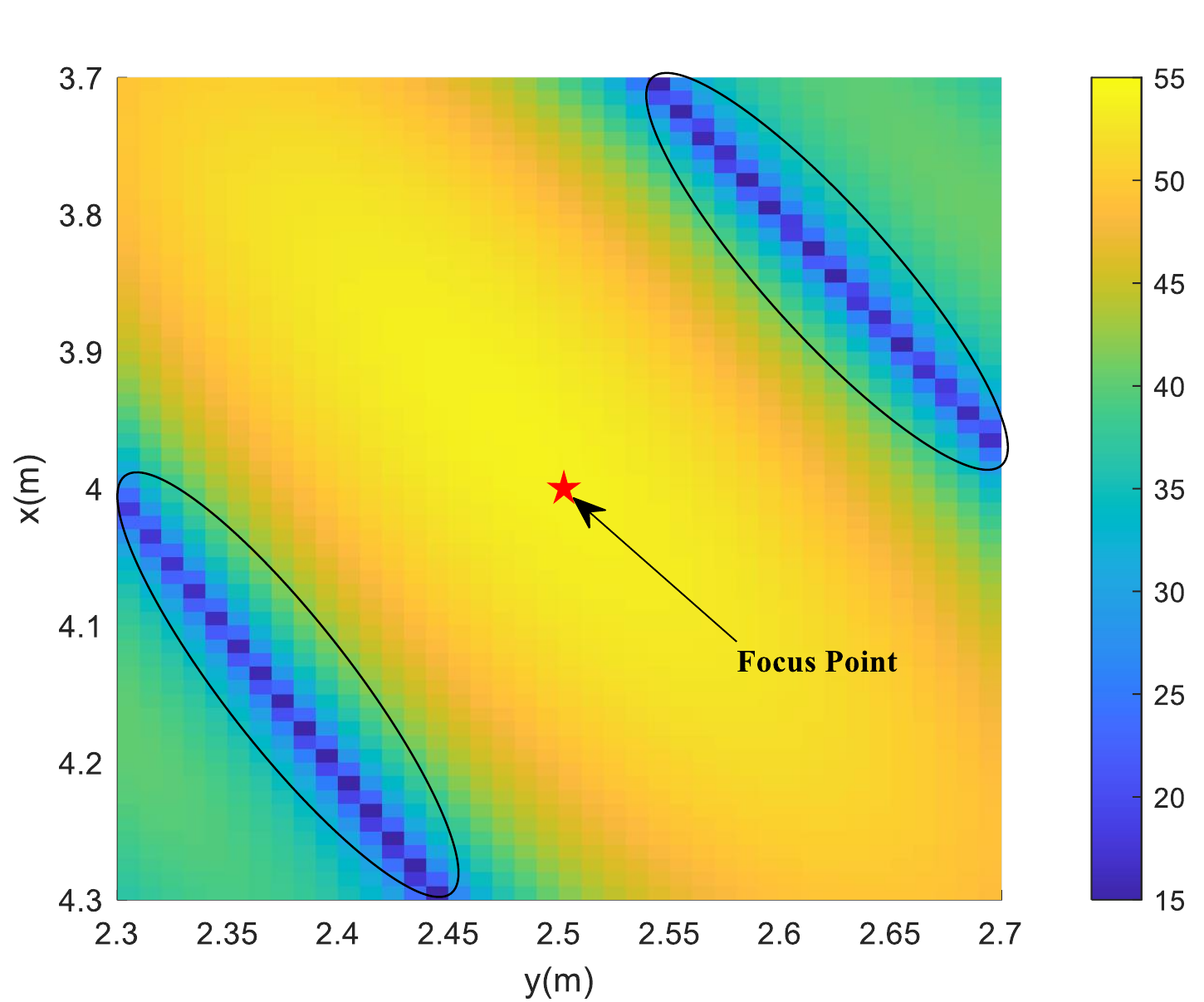}
	} 
    \hspace{-0.8cm}
    \caption{The PEB and the received SNR at $z=-1$ plane with focusing control scheme at the RIS. (a) The PEB of asynchronous scenario. Bandwidth $=40$MHz. (b) The PEB of synchronous scenario. (c) The received SNR.}\label{Fig.focus1}
\end{figure*}
% \begin{figure}[!h] 
%     \centering 
%     \hspace{-0.3cm}    
%     \includegraphics[width=9cm]{simu_SNR_focus.pdf}
%     \caption{}
% \end{figure}
\begin{figure}
    \centering
    \color{blue}
    \hspace{-0.5cm}
	\subfloat[]{
		\includegraphics[width=4cm]{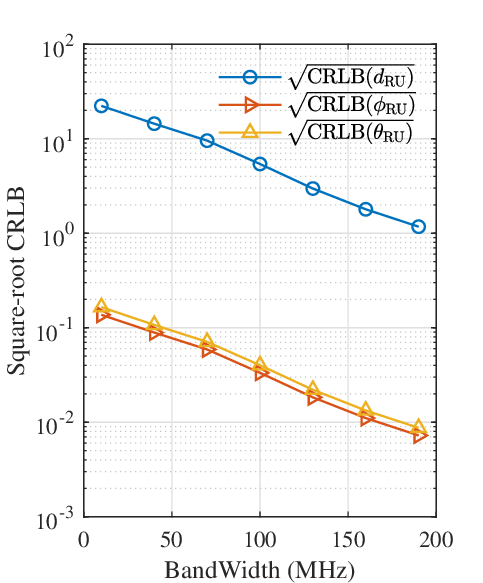}
	}
	\hspace{-0.5cm}
	\subfloat[]{
		\includegraphics[width=4cm]{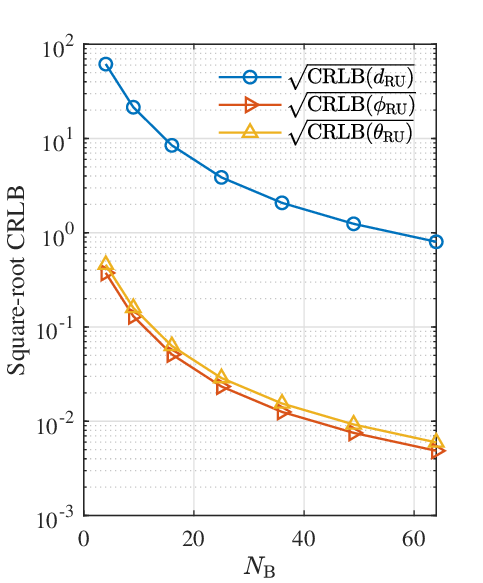}
	} 
    \hspace{-0.5cm}
    \caption{The square-root CRLBs of the intermediate parameters at the focus point as a function of bandwidth and $N_\mathrm{B}$ in the asynchronous scenario. (a) $N_\mathrm{B}=4$ is fixed. (b) Bandwidth $=10$ MHz is fixed.}\label{Fig.focus2}
\end{figure}
\begin{figure}[!t] 
    \centering 
    \hspace{-0.3cm} 
    \color{blue}   
    \includegraphics[width=7.5cm]{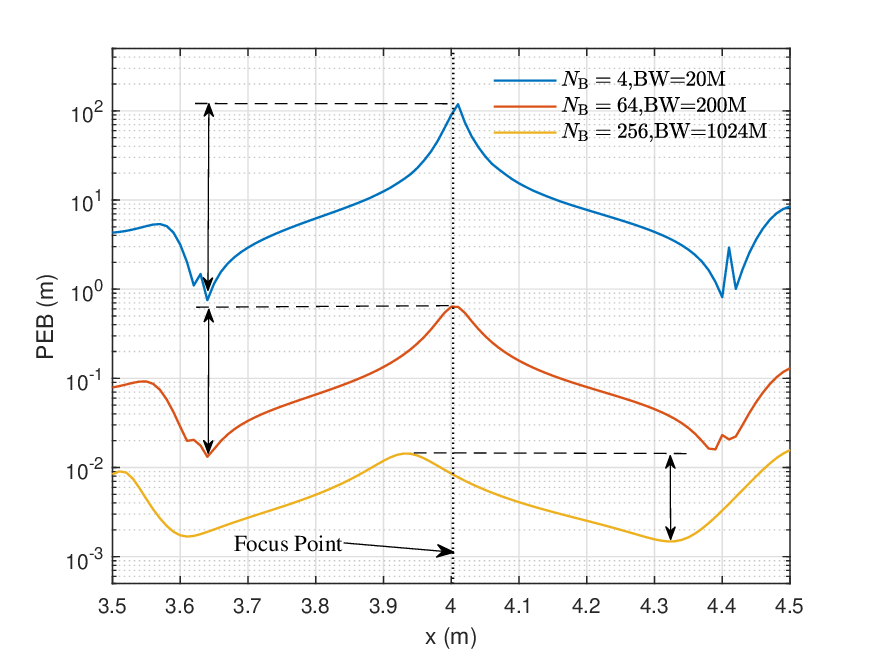}
    \caption{The PEBs along $x$ axis with focusing control scheme under different system configurations. $y=2.5$ and $z=-1$ are fixed. }\label{Fig.focus3}
\end{figure}
\begin{figure*}[!t] 
    \centering 
    \color{blue}
    \hspace{-0.8cm}       
    \subfloat[]{
		\includegraphics[width=5.5cm]{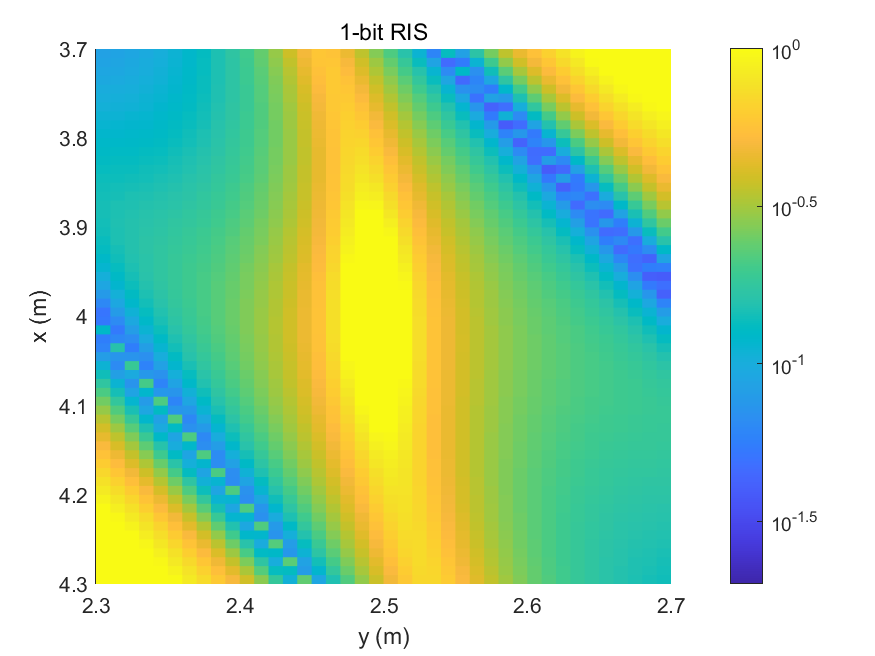}
	}    
	\hspace{-0.3cm}
    \subfloat[]{
		\includegraphics[width=5.5cm]{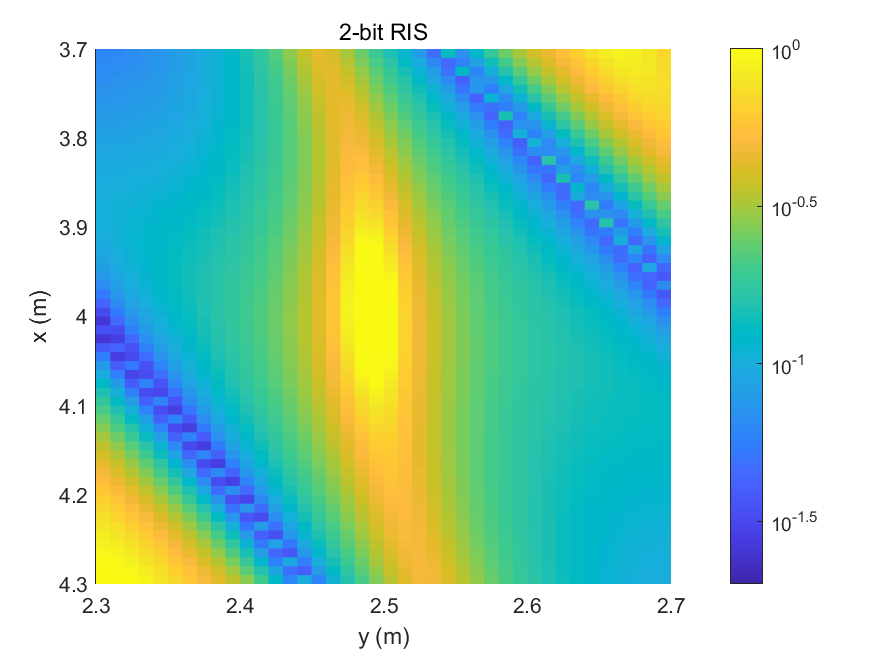}
	}
	\hspace{-0.3cm}
	\subfloat[]{
		\includegraphics[width=5.5cm]{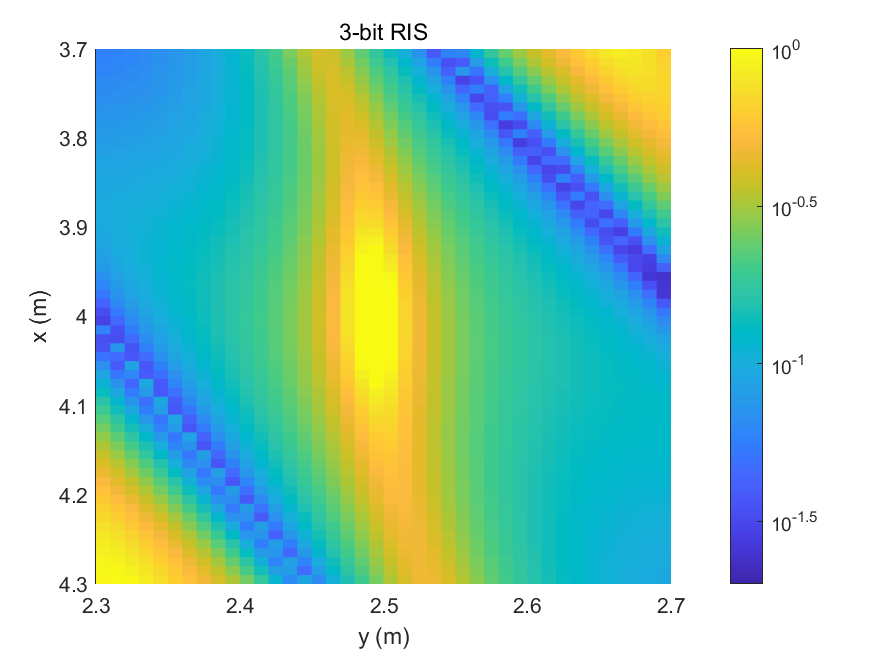}
	} 
    \hspace{-0.8cm}
    \caption{The PEB at $z=-1$ plane when adopting focusing control scheme with discrete RIS coefficients.} \label{Fig.quanti}
\end{figure*}
\begin{figure*}[!t] 
    \centering 
    \color{blue}
    \hspace{-0.8cm}       
    \subfloat[]{
		\includegraphics[width=5.5cm]{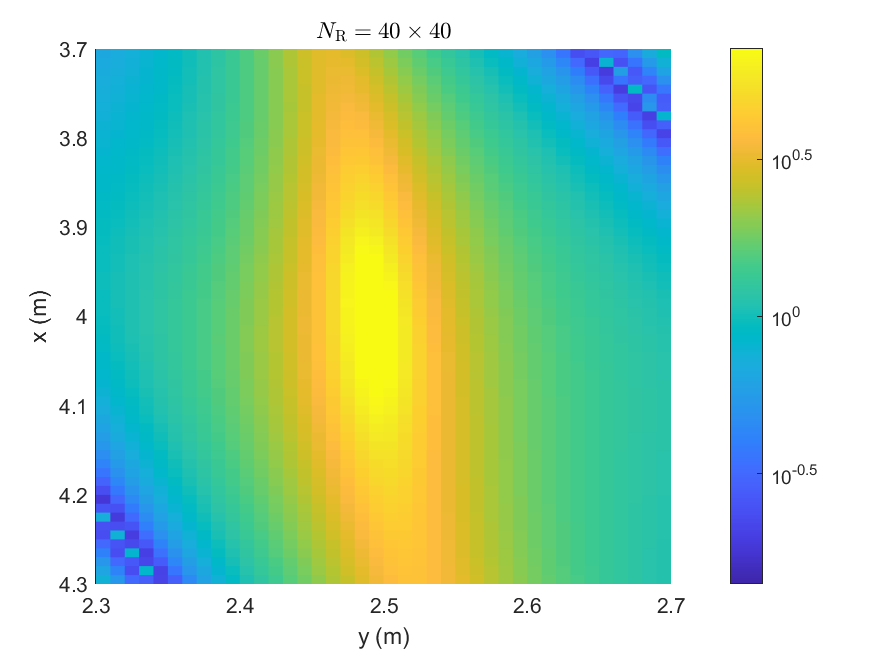}
	}    
	\hspace{-0.3cm}
    \subfloat[]{
		\includegraphics[width=5.5cm]{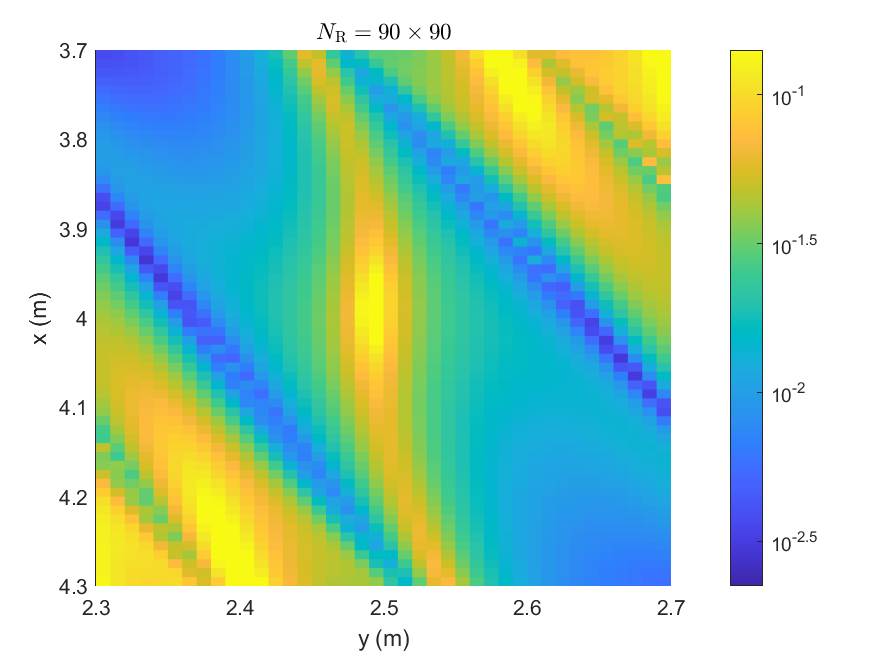}
	}
    \caption{The PEB at $z=-1$ plane when adopting focusing control scheme with different RIS sizes.}\label{Fig.RISsize}
\end{figure*}

\subsection{Evaluation of the Focusing Control Scheme}\label{sec:V-C}
In this subsection, the PEB performances are evaluated when the focusing control plan is adopted for the RIS in both synchronous and asynchronous scenarios. The RIS is set to focus the signal with frequency 28 GHz (carrier frequency) from position $(4, 2.5, -1)$ to the reference point of the BS antenna array in a single time slot. We get the samples of the received signal from all the antennas and sub-carriers in this time slot. 

Figs. \ref{Fig.focus1}(a) and \ref{Fig.focus1}(b) depict the PEBs at $z=-1$ plane with focusing scheme in asynchronous and synchronous scenarios, respectively. The corresponding received SNR of this region is depicted in Fig. \ref{Fig.focus1}(c). The bandwidth of the transmitted signal is set as $40$ MHz. From Fig. \ref{Fig.focus1}(a), a `peak' of the PEB occurs around the focus point (as circled by the red ellipse), which indicates a worse localization performance in this region. However, the region around the focus point is with the highest SNR as confirmed in Fig. \ref{Fig.focus1}(c). Besides, Comparing the regions circled by the black ellipses in Figs. \ref{Fig.focus1}(a) and (c), the received SNRs of the two regions are relatively much lower than the area around the focus point, but the PEB performances are much better. \textcolor{blue}{The above results also indicate that, if we intend to minimize the PEB at a specific position $\mathbf{p}_0$, a good choice for RIS coefficients that may achieve a decent PEB performance is to try focusing the signal \emph{near} $\mathbf{p}_0$, instead of precisely \emph{at} this position. Based on this observation, we are able to efficiently obtain a good initial point for further optimizing the RIS coefficients.} Note that although in this subsection, the received signals are not perfectly aligned at most of the antennas and sub-carriers as shown in Figs. \ref{Fig.focuscases}(e) and (f), and the EFIM for $\boldsymbol{\eta}$ is not $\mathbf{0}$ at the focus point (unlike the Case 1 and Case 2 in Proposition \ref{prop.4}), the simulation results still verify that in the asynchronous scenario, the localization performance may be degraded with the increase of the received SNR, which is consistent with the discussion in Remark \ref{remark3}. \textcolor{blue}{Moreover, the simulation results also verify that the discussion in Section \ref{sec:IV-C} still holds in the scenario that the multi-paths between the BS and the RIS cannot be ignored.} However, from Fig. \ref{Fig.focus1}(b), in the synchronous scenario, the `peak' of the PEB around the focus point disappears and in general the PEB decreases with the increase of the received SNR, which is unlike the asynchronous scenario. That is because, firstly, in the synchronous scenario the TOA information is attainable, thus the distance parameter $d_\mathrm{RU}$ can be directly recovered from the TOA which does not rely on the phase differences at different RIS elements. Therefore, the higher SNR will generally lead to better distance estimation performance. Besides, in Fig. \ref{Fig.focus2}, the square-root CRLBs vs. the bandwidth and $N_\mathrm{B}$ at the focus point are depicted when adopting the focusing scheme. Although the direction of the UE still relies on the phase-differences at the RIS elements in the synchronous scenario, from Fig. \ref{Fig.focus2}, the high-accuracy direction related parameters (namely $\theta_\mathrm{RU}$ and $\phi_\mathrm{RU}$) can be obtained much more easily than the distance parameters when only relying on the phase differences caused by the wavefront. For example, note that $0.01$ rad (less than $0.6^\circ$) estimation error of $\theta_\mathrm{RU}$ and $\phi_\mathrm{RU}$ only leads to less than $0.1$ metre localization error when $d_\mathrm{RU}$ is less than $10$ metres (assuming $d_\mathrm{RU}$ is accurately estimated). However, from Fig. \ref{Fig.focus2}, when about $0.01$ rad root-square CRLBs of $\theta_\mathrm{RU}$ and $\phi_\mathrm{RU}$ are achieved, the square-root CRLB of $d_\mathrm{RU}$ is more than 1 metre, which means that the estimation error of $d_\mathrm{RU}$ is dominated in this scenario.
% when the square-root CRLB of $d_\mathrm{RU}$ is more than 1 metre, less than $0.6^\circ$ (about $0.01$ rad) root-square CRLBs of $\theta_\mathrm{RU}$ and $\phi_\mathrm{RU}$ are attainable, which only leads to less than $0.1$ metre localization error when $d_\mathrm{RU}$ is less than $10$ metres in the perspective view of CRLB. 
In other words, obtaining high-accuracy distance parameter is the bottleneck of the near-field high-accuracy localization. Therefore, in the synchronous scenario where the higher SNR could lead to better $d_\mathrm{RU}$ estimation performance, in general the PEB will also be improved with the increase of SNR. Moreover, comparing Fig. \ref{Fig.focus2}(a) with Fig. \ref{Fig.focus2}(b), the estimation performance of the intermediate parameters can be improved by either increasing the spatial resource (namely $N_\mathrm{B}$) or the bandwidth resource, which is consistent with the discussion in Remark \ref{remark1}.

In Fig. \ref{Fig.focus3}, the PEBs along $x$ axis are depicted, and $y=2.5$ and $z=-1$ are fixed to evaluate the influence of different system configurations on the relative size of the `peak' around the focus point\footnote{The relative sizes of the `peaks' under different system configurations are comparable when using logarithmic scale on $y$ axis for PEB.}. From Fig. \ref{Fig.focus3}, the `peak' of the PEB will be alleviated when the number of the BS antennas and the bandwidth increase. That is because the signal reflected from the RIS will tend to be less aligned with the increase of antennas and bandwidth, which leads to more significant phase differences at the RIS from the focus point. 

\textcolor{blue}{In Fig. \ref{Fig.quanti}, we investigate the impact of quantizing the phase of the RIS, which can reduce the complexity of the RIS. Comparing with Fig. \ref{Fig.focus1}(a), the differences on PEB are negligible when the 3-bit RIS is adopted. Besides, in Fig. \ref{Fig.RISsize}, we evaluate the PEB performances for various RIS sizes. As expected, the RIS with a larger size could generally lead to a better PEB performance, and the `peak' area around the focus point shrinks with the increase of the RIS elements. In other words, increase the size of the RIS will alleviate the PEB performance degradation effect around the focus point.}

\section{Conclusions}\label{sec:conclusion}
In this paper, the PEB of the RIS-based asynchronous localization and the EFI for the position-related intermediate parameters have been derived \textcolor{blue}{under the circumstances that the multi-paths between the BS and the RIS are considered. The more accurate model is adopted that takes the amplitude differences across the RIS and the BS into account.} Based on the derived EFI, we first prove that in the asynchronous scenario, it is theoretically possible to localize the UE when the near-field spherical wavefront is considered in the RIS-UE part of the channel. However, with the increase of the distance from the UE to the RIS, the EFI for the distance parameter tends to $0$. We then revealed that when the near-field model was considered for the BS-RIS part of the channel, the multiple antennas at the BS can provide independent spatial gain, while only the power gain is provided by the antennas when this part of the channel works in the far-field scenario. The spatial gain outperforms the power gain when there is no much information about the UE position at each single antenna. But when there is adequate information at each antenna, the probability that the power gain achieves better PEB performance will increase. 
We also showed that the well-known focusing control scheme for the RIS, which maximizes the received SNR, is not always a good choice for localization, since it may reduce the phase differences caused by the spherical wavefront at the RIS and degrade the localization performance in the asynchronous scenario.
% The performance of the well-known focusing control scheme for RIS which maximizes the received SNR has been investigated. However, we find that it is not always a good choice to take maximizing the SNR as the design criterion for RIS coefficients since it may degrade the localization performance in the asynchronous scenario. 
% The comprehensive analysis in this paper provides useful insights in the design of RIS-assisted localization system. 

\appendices
\renewcommand{\theequation}{\thesection.\arabic{equation}}
\setcounter{equation}{0}
\section{} \label{appendix1}
In this appendix we prove Proposition \ref{prop.1}. When the far-field scenario is considered, according to (\refeq{drU_far}) we have $\frac{\partial {d}_{r\mathrm{U}}}{\partial d_{\mathrm{RU}}} = 1$ for all $r$. While in the near-field scenario, when we let $d_\mathrm{RU}\to\infty$, combining (\refeq{dru_near}) and (\refeq{DdRU_eta_near_1}) we also have
\begin{align}
    \label{DdrU_dRU_far}
    \lim_{d_\mathrm{RU}\to\infty}\frac{\partial {d}_{r\mathrm{U}}}{\partial d_{\mathrm{RU}}} =\lim_{d_\mathrm{RU}\to\infty}\frac{1}{d_{r\mathrm{U}}}\left( d_{\mathrm{RU}}+\Gamma _{r\mathrm{U}} \right)=1 .  
\end{align}
Therefore, in the above two cases one obtains $\frac{\partial \mathbf{d}_{\mathrm{RU}}}{\partial \left[ \boldsymbol{\eta } \right] _3}=\frac{\partial \mathbf{d}_{\mathrm{RU}}}{\partial d_\mathrm{RU}}=\mathbf{1}_{N_\mathrm{R}}$ and \textcolor{blue}{$\boldsymbol{\gamma}_{\mathrm{RU},n}\to\frac{\sqrt{2P_{\mathrm{t}}}\lambda _n}{4\pi d_{\mathrm{RU}}}\mathbf{1}$. Under these circumstances we have
\begin{align}
    \label{hbnfar}
    \frac{\sqrt{2P_{\mathrm{t}}}\lambda _n}{4\pi}\dot{\mathbf{h}}_{b,n}=-\frac{1}{d_{\mathrm{RU}}}\tilde{\mathbf{h}}_{b,n}.
\end{align}
Combining (\refeq{hbnfar}), each element of $\bar{\mathbf{J}}_{\boldsymbol{\mu}}$ can be further calculated as 
\begin{subequations}
    \label{far_barJ}
    \begin{align}
        &[\bar{\mathbf{J}}_{\boldsymbol{\mu }}]_{1,1}=\sum_{t=1}^T{\sum_{n=1}^N{\sum_{b=1}^{N_{\mathrm{B}}}{\frac{2|x_{n,t}|^2}{\sigma ^2}\left| \boldsymbol{\phi }_{t}^{\top}\tilde{\mathbf{h}}_{b,n} \right|^2}}}, \quad [\bar{\mathbf{J}}_{\boldsymbol{\mu }}]_{1,2}= [\bar{\mathbf{J}}_{\boldsymbol{\mu }}]_{2,1}=0,\\
       & [\bar{\mathbf{J}}_{\boldsymbol{\mu }}]_{1,3}=[\bar{\mathbf{J}}_{\boldsymbol{\mu }}]_{3,1}=\sum_{t=1}^T{\sum_{n=1}^N{\sum_{b=1}^{N_{\mathrm{B}}}{-\frac{2\alpha |x_{n,t}|^2}{d_{\mathrm{RU}}\sigma ^2}\left| \boldsymbol{\phi }_{t}^{\top}\tilde{\mathbf{h}}_{b,n} \right|^2}}}, \\
        &[\bar{\mathbf{J}}_{\boldsymbol{\mu }}]_{1,k+2}=[\bar{\mathbf{J}}_{\boldsymbol{\mu }}]_{k+2,1}=\sum_{t=1}^T{\sum_{n=1}^N{\sum_{b=1}^{N_{\mathrm{B}}}{\Re \biggl\{ \frac{-j4\pi f_n\alpha |x_n|^2}{c\sigma ^2}\tilde{\mathbf{h}}_{b,n}^{\mathrm{H}}\boldsymbol{\phi }_{t}^{*}\boldsymbol{\phi }_{t}^{\top}\tilde{\mathbf{d}}_{\boldsymbol{\eta }_k} \biggr\}}}} \\
        &[\bar{\mathbf{J}}_{\boldsymbol{\mu }}]_{2,2}=[\bar{\mathbf{J}}_{\boldsymbol{\mu }}]_{2,3}=[\bar{\mathbf{J}}_{\boldsymbol{\mu }}]_{3,2}=\sum_{t=1}^T{\sum_{n=1}^N{\sum_{b=1}^{N_{\mathrm{B}}}{\frac{8\pi ^2\alpha ^2f_{n}^{2}|x_n|^2}{c^2\sigma ^2}\left| \boldsymbol{\phi }_{t}^{\top}\tilde{\mathbf{h}}_{b,n} \right|^2}}} ,\\
        &[\bar{\mathbf{J}}_{\boldsymbol{\mu }}]_{2,k+2}=[\bar{\mathbf{J}}_{\boldsymbol{\mu }}]_{k+2,2}=\sum_{t=1}^T{\sum_{n=1}^N{\sum_{b=1}^{N_{\mathrm{B}}}{\Re \biggl\{ \frac{8\pi ^2\alpha ^2f_{n}^{2}|x_n|^2}{c^2\sigma ^2}\tilde{\mathbf{h}}_{b,n}^{\mathrm{H}}\boldsymbol{\phi }_{t}^{*}\boldsymbol{\phi }_{t}^{\top}\tilde{\mathbf{d}}_{\boldsymbol{\eta }_k} \biggr\}}}} ,\\
        &[\bar{\mathbf{J}}_{\boldsymbol{\mu }}]_{3,3}=\sum_{t=1}^T{\sum_{n=1}^N{\sum_{b=1}^{N_{\mathrm{B}}}{\left( \frac{8\pi ^2\alpha ^2f_{n}^{2}|x_n|^2}{c^2\sigma ^2}+\frac{2\alpha ^2|x_n|^2}{d_{\mathrm{RU}}^{2}\sigma ^2} \right) \left| \boldsymbol{\phi }_{t}^{\top}\tilde{\mathbf{h}}_{b,n} \right|^2}}}, \\
        &[\bar{\mathbf{J}}_{\boldsymbol{\mu }}]_{k+2,3}=[\bar{\mathbf{J}}_{\boldsymbol{\mu }}]_{3,k+2}=\sum_{t=1}^T{\sum_{n=1}^N{\sum_{b=1}^{N_{\mathrm{B}}}{\Re \biggl\{ \left( \frac{8\pi ^2\alpha ^2f_{n}^{2}|x_n|^2}{c^2\sigma ^2}+\frac{j4\pi f_n\alpha ^2|x_n|^2}{d_{\mathrm{RU}}c\sigma ^2} \right) \tilde{\mathbf{h}}_{b,n}^{\mathrm{H}}\boldsymbol{\phi }_{t}^{*}\boldsymbol{\phi }_{t}^{\top}\tilde{\mathbf{d}}_{\boldsymbol{\eta }_k} \biggr\}}}}.
    \end{align} 
\end{subequations}
where $k\in \{ 2,3 \}$. Besides, since in the far-field scenario $\left\| \boldsymbol{\gamma }_{\mathrm{RU},n} \right\| ^2=N_{\mathrm{R}}\left( \frac{\sqrt{2P_{\mathrm{t}}}\lambda _n}{4\pi d_{\mathrm{RU}}} \right) ^2\triangleq N_{\mathrm{R}}\gamma _{\mathrm{RU},n}^{2}$, 
% \begin{align}
%     \left\| \boldsymbol{\gamma }_{\mathrm{RU},n} \right\| ^2=N_{\mathrm{R}}\left( \frac{\sqrt{2P_{\mathrm{t}}}\lambda _n}{4\pi d_{\mathrm{RU}}} \right) ^2\triangleq N_{\mathrm{R}}\gamma _{\mathrm{RU},n}^{2}, 
% \end{align}
we have
\begin{subequations}
    \label{farDC}
    \begin{align}
        &\frac{\partial \mathbf{C}_{\tilde{\mathbf{w}}_{n,t}}}{\partial \alpha}=2N_{\mathrm{R}}\alpha |x_{n,t}|^2N_{\mathrm{R}}\gamma _{\mathrm{RU},n}^{2}\sigma _{\mathrm{H}}^{2}\mathbf{I}\triangleq {c}_{\alpha \left( n,t \right)}\sigma _{\mathrm{H}}^{2}\mathbf{I}, \\
        &\frac{\partial \mathbf{C}_{\tilde{\mathbf{w}}_{n,t}}}{\partial d_{\mathrm{RU}}}=-\alpha ^2|x_{n,t}|^2\gamma _{\mathrm{RU},n}\frac{\sqrt{2P_{\mathrm{t}}}\lambda _n}{2\pi d_{\mathrm{RU}}^{2}}\sigma _{\mathrm{H}}^{2}\mathbf{I}\triangleq c_{d\left( n,t \right)}\sigma _{\mathrm{H}}^{2}\mathbf{I}, \\
        &\frac{\partial \mathbf{C}_{\tilde{\mathbf{w}}_{n,t}}}{\partial \left( c\xi \right)}=\frac{\partial \mathbf{C}_{\tilde{\mathbf{w}}_{n,t}}}{\partial \varphi _{\mathrm{RU}}}=\frac{\partial \mathbf{C}_{\tilde{\mathbf{w}}_{n,t}}}{\partial \theta _{\mathrm{RU}}}=\mathbf{0}.
    \end{align}
\end{subequations}
Note that from (\refeq{farDC}), 
\begin{align}  
    &c_{d\left( n,t \right)}=-\frac{\alpha}{d_{\mathrm{RU}}}c_{\alpha \left( n,t \right)}, \label{far_c_relation}\\
    &[\bar{\mathbf{J}}_{\tilde{\mathbf{w}}_{n,t}}]_{:,2}=\left( [\bar{\mathbf{J}}_{\tilde{\mathbf{w}}_{n,t}}]_{2,:} \right) ^{\top}=\mathbf{0}, \quad [\bar{\mathbf{J}}_{\tilde{\mathbf{w}}_{n,t}}]_{4:5,:}=\left( [\bar{\mathbf{J}}_{\tilde{\mathbf{w}}_{n,t}}]_{:,4:5} \right) ^{\top}=\mathbf{0} \label{Jwfar0}
\end{align}
Thus according to (\refeq{far_barJ}) and (\refeq{farDC})-(\refeq{far_c_relation}), it can be verified that  
\begin{align}
&-\frac{\alpha}{d_{\mathrm{RU}}}[\bar{\mathbf{J}}_{\boldsymbol{\mu }}]_{\left( 1:5 \right) \setminus 3,1}+[\bar{\mathbf{J}}_{\boldsymbol{\mu }}]_{\left( 1:5 \right) \setminus 3,2}=[\bar{\mathbf{J}}_{\boldsymbol{\mu }}]_{\left( 1:5 \right) \setminus 3,3} \\
&-\frac{\alpha}{d_{\mathrm{RU}}}[\bar{\mathbf{J}}_{\tilde{\mathbf{w}}_{n,t}}]_{1,1}=[\bar{\mathbf{J}}_{\tilde{\mathbf{w}}_{n,t}}]_{1,3}, \quad -\frac{\alpha}{d_{\mathrm{RU}}}[\bar{\mathbf{J}}_{\tilde{\mathbf{w}}_{n,t}}]_{3,1}=[\bar{\mathbf{J}}_{\tilde{\mathbf{w}}_{n,t}}]_{3,3} \label{Jwrelation}
\end{align}
Therefore, combining (\refeq{EFI}) and (\refeq{Jwfar0})-(\refeq{Jwrelation}), regardless of whether $\bar{\mathbf{J}}_{\tilde{\mathbf{w}}}=\mathbf{0}$\footnote{\textcolor{blue}{Which corresponds to the case that the multi-paths between the BS and the RIS are not considered.}}, we have 
\begin{align}
\bar{J}_\mathrm{E}(d_\mathrm{RU})&=
% [\bar{\boldsymbol{\mathcal{J}}}]_{3,3}-[\bar{\boldsymbol{\mathcal{J}}}]_{3,\left( 1:5 \right) \setminus 3}[\bar{\boldsymbol{\mathcal{J}}}]_{\left( 1:5 \right) \setminus 3,\left( 1:5 \right) \setminus 3}^{-1}[\bar{\boldsymbol{\mathcal{J}}}]_{\left( 1:5 \right) \setminus 3,3} \nonumber \\
[\bar{\boldsymbol{\mathcal{J}}}]_{3,3}-[\bar{\boldsymbol{\mathcal{J}}}]_{3,\left( 1:5 \right) \setminus 3}\cdot \left[ -\alpha /d_{\mathrm{RU}},1,0,0 \right] ^{\top} \nonumber \\
&=[\bar{\boldsymbol{\mathcal{J}}}]_{3,3}-(-\frac{\alpha}{d_{\mathrm{RU}}}[\bar{\boldsymbol{\mathcal{J}}}]_{3,1}+[\bar{\boldsymbol{\mathcal{J}}}]_{3,2})=0.
\end{align}
Thus complete the proof.}

\setcounter{equation}{0}
\section{} \label{appendix2}
In this appendix we prove Proposition \ref{prop.2}. We first rewrite $\mu_{b,n,t}$ in the far-field scenario as 
\begin{align}
    \mu_{b,n,t}&\color{blue}=\alpha x_ne^{-j2\pi f_n\xi}\sum_{r=1}^{N_{\mathrm{R}}}{\frac{\lambda _n}{4\pi d_{\mathrm{B}r}}\frac{\sqrt{2P_{\mathrm{t}}}\lambda _n}{4\pi d_{r\mathrm{U}}}e^{j\phi _{r,t}}e^{-j2\pi f_n\left( d{ _{br}}+d_{r\mathrm{U}} \right) /c}} \\
    &\simeq e^{-j2\pi f_n\left( d_{\mathrm{BR}}+\Gamma _{\mathrm{B},b} \right) /c} \cdot F_{n,t} \label{mu_BS-RIS_far},
\end{align}
where $\phi_{r,t}=[\boldsymbol{\phi}_t]_r$; $F_{n,t}$ is defined as 
\begin{align}
    \color{blue}
    F_{n,t}\triangleq \alpha x_ne^{-j2\pi f_n\xi}\sum_{r=1}^{N_{\mathrm{R}}}{\frac{\lambda _n}{4\pi d_{\mathrm{B}r}}\frac{\sqrt{2P_{\mathrm{t}}}\lambda _n}{4\pi d_{r\mathrm{U}}}e^{j\phi _{r,t}}e^{-j2\pi f_n\left( \Gamma _{r\mathrm{B}}+d_{r\mathrm{U}} \right) /c}}.
\end{align}
Eq.(\refeq{mu_BS-RIS_far}) holds when the BS-RIS part of the channel works in the far-field scenario (by combining (\refeq{HBR_far}) and (\refeq{aB_aRB})). It is straightforward to verify that $\frac{\partial e^{-j2\pi f_n\left( d_{\mathrm{BR}}+\Gamma _{\mathrm{B},b} \right) /c}}{\partial \mathbf{\Theta }}=\mathbf{0}$.
% \begin{align}
%     \frac{\partial e^{-j2\pi f_n\left( d_{\mathrm{BR}}+\Gamma _{\mathrm{B},b} \right) /c}}{\partial \mathbf{\Theta }}=\mathbf{0}.
% \end{align}
Therefore, in this case we have $\frac{\partial \mu _{b,n,t}}{\partial \mathbf{\Theta }^{\top}}=e^{-j2\pi f_n\left( d_{\mathrm{BR}}+\Gamma _{\mathrm{B},b} \right) /c}\frac{\partial F_{n,t}}{\partial \mathbf{\Theta }^{\top}}$,
and each element of \textcolor{blue}{$\mathbf{J}^{(\mathrm{B})}_{\boldsymbol{\mu}b}$} is then given by
\begin{align}
    &[ \mathbf{J}^{(\mathrm{B})}_{\boldsymbol{\mu}b} ] _{i,j}=\sum_{t=1}^T{\sum_{n=1}^N{\frac{2}{\sigma ^2}\Re \left\{ \frac{\partial \mu _{b,n,t}^{*}}{\partial \left[ \mathbf{\Theta } \right] _i}\frac{\partial \mu _{b,n,t}}{\partial \left[ \mathbf{\Theta } \right] _j} \right\}}} \nonumber
    \\
    &=\sum_{t=1}^T{\sum_{n=1}^N{\frac{2}{\sigma ^2}\Re \left\{ \left| e^{-j2\pi f_n\left( d_{\mathrm{BR}}+\Gamma _{\mathrm{B},b} \right) /c} \right|^2\frac{\partial F_{n,t}^{*}}{\partial \left[ \mathbf{\Theta } \right] _i}\frac{\partial F_{n,t}}{\partial \left[ \mathbf{\Theta } \right] _j} \right\}}} =\sum_{t=1}^T{\sum_{n=1}^N{\frac{2}{\sigma ^2}\Re \left\{ \frac{\partial F_{n,t}^{*}}{\partial \left[ \mathbf{\Theta } \right] _i}\frac{\partial F_{n,t}}{\partial \left[ \mathbf{\Theta } \right] _j} \right\}}},\mspace{-3mu}\label{HBRfarFIMb}
\end{align}
where $1\le i,j\le 5$. From (\refeq{HBRfarFIMb}), the expression of $\mathbf{J}^{(\mathrm{B})}_{\boldsymbol{\mu}b}$ is not related to variable $b$. Therefore,  $\mathbf{J}^{(\mathrm{B})}_{\boldsymbol{\mu}b}$ is identical for all $b$  when the far-field scenario is considered for \textcolor{blue}{$\mathbf{H}_{\mathrm{BR},n}^{(\mathrm{d})}$}. \textcolor{blue}{Thus we have $\mathbf{J}_{\boldsymbol{\mu}}\propto N_\mathrm{B}$. Besides, from (\refeq{Jw1}), $\mathbf{J}_{\tilde{\mathbf{w}}}\propto N_\mathrm{B}$. Therefore, $\boldsymbol{\mathcal{J}}=\mathbf{J}_{\boldsymbol{\mu}}+\mathbf{J}_{\tilde{\mathbf{w}}}\propto N_\mathrm{B}$ and the Proposition \ref{prop.2} is then proved.}

% \vspace{-0.5cm}
\setcounter{equation}{0}
\textcolor{blue}{
\section{} \label{appendix3}
In this appendix we prove Proposition \ref{prop.4}. Note that when the multi-paths effect is ignored, we have $\boldsymbol{\mathcal{J}} =\mathbf{J}_{\boldsymbol{\mu }}$.
    Denote the FIM for $\bar{\mathbf{\Theta}}$ in Case 1 as $\mathbf{\bar{J}}_{\mathrm{C}1}$, one obtains $\mathbf{\bar{J}}_{\mathrm{C}1}=\sum_{t_1=1}^{N_{\mathrm{B}}}{\mathbf{J}_{t_1,n_0,t_1}}$.
    % \begin{align}
    %     \mathbf{\bar{J}}_{\mathrm{C}1}=\sum_{t_1=1}^{N_{\mathrm{B}}}{\mathbf{J}_{t_1,n_0,t_1}}.
    % \end{align}
    When $\boldsymbol{\phi}_{t_1}=(\mathring{\mathbf{h}}_{t_1\mathrm{R},n_0}\circledast \mathring{\mathbf{h}}_{\mathrm{RU},n_0})^*$, combining (\refeq{Jbnt2}), (\refeq{Jbarbnt1}) and (\refeq{Jbarbnt_ele}), each element of $\mathbf{\bar{J}}_{\mathrm{C}1}$ is then calculated as
    \begin{subequations}
        \label{JC1}
        \begin{align}
            &[\mathbf{\bar{J}}_{\mathrm{C}1}]_{1,1}=\frac{2|x_{n_0}|^2}{\sigma ^2}\sum_{t_1=1}^{N_{\mathrm{B}}}{\left( \boldsymbol{\gamma }_{t_1\mathrm{R},n_0}^{\top}\boldsymbol{\gamma }_{\mathrm{RU},n_0} \right) ^2}\approx \frac{2|x_{n_0}|^2}{\sigma ^2}N_{\mathrm{B}}\left( \boldsymbol{\gamma }_{\mathrm{BR},n_0}^{\top}\boldsymbol{\gamma }_{\mathrm{RU},n_0} \right) ^2,\\
            &[\mathbf{\bar{J}}_{\mathrm{C}1}]_{1,2}=[\mathbf{\bar{J}}_{\mathrm{C}1}]_{2,1}=0, \label{JC1_12} \\
            &[\mathbf{\bar{J}}_{\mathrm{C}1}]_{1,k}=[\mathbf{\bar{J}}_{\mathrm{C}1}]_{k,1} =\sum_{t_1=1}^{N_{\mathrm{B}}}{\Re \biggl\{ \frac{-j4\pi \alpha f_{n_0}|x_{n_0}|^2}{c\sigma ^2}\left( \boldsymbol{\gamma }_{t_1\mathrm{R},n_0}^{\top}\boldsymbol{\gamma }_{\mathrm{RU},n_0} \right) \left( \boldsymbol{\gamma }_{t_1\mathrm{R},n_0}\circledast \boldsymbol{\gamma }_{\mathrm{RU},n_0} \right) ^{\top}\frac{\partial \mathbf{d}_{\mathrm{RU}}}{\partial [\bar{\mathbf{\Theta}}]_k} \biggr\}} \nonumber
            \\
            &\mspace{180mu}   +\Re \biggl\{ \frac{2\sqrt{2P_{\mathrm{t}}}\alpha \lambda _n|x_{n_0}|^2}{4\pi \sigma ^2}\left( \boldsymbol{\gamma }_{t_1\mathrm{R},n_0}^{\top}\boldsymbol{\gamma }_{\mathrm{RU},n_0} \right) \left( \boldsymbol{\gamma }_{t_1\mathrm{R},n_0}\circledast \dot{\mathbf{d}}_{\mathrm{RU}} \right) ^{\top}\frac{\partial \mathbf{d}_{\mathrm{RU}}}{\partial [\bar{\mathbf{\Theta}}]_k} \biggr\} \nonumber \\
            % &=\frac{2\sqrt{2P_{\mathrm{t}}}\alpha \lambda _n|x_{n_0}|^2}{4\pi \sigma ^2}\sum_{t_1=1}^{N_{\mathrm{B}}}{\left( \boldsymbol{\gamma }_{t_1\mathrm{R},n_0}^{\top}\boldsymbol{\gamma }_{\mathrm{RU},n_0} \right) \left( \boldsymbol{\gamma }_{t_1\mathrm{R},n_0}\circledast \dot{\mathbf{d}}_{\mathrm{RU}} \right) ^{\top}\frac{\partial \mathbf{d}_{\mathrm{RU}}}{\partial [\bar{\mathbf{\Theta}}]_k}}, \\
            &\mspace{135mu}\approx  \frac{2\sqrt{2P_{\mathrm{t}}}\alpha \lambda _n|x_{n_0}|^2}{4\pi \sigma ^2}N_{\mathrm{B}}\left( \boldsymbol{\gamma }_{\mathrm{BR},n_0}^{\top}\boldsymbol{\gamma }_{\mathrm{RU},n_0} \right) \left( \boldsymbol{\gamma }_{\mathrm{BR},n_0}\circledast \dot{\mathbf{d}}_{\mathrm{RU}} \right) ^{\top}\frac{\partial \mathbf{d}_{\mathrm{RU}}}{\partial [\bar{\mathbf{\Theta}}]_k} .     \label{JC1_c}\\
            &[\mathbf{\bar{J}}_{\mathrm{C}1}]_{2,2}\approx \frac{8\pi ^2f_{n_0}^{2}\alpha ^2\left| x_{n_0} \right|^2}{c^2\sigma ^2}N_{\mathrm{B}}\left( \boldsymbol{\gamma }_{\mathrm{BR},n_0}^{\top}\boldsymbol{\gamma }_{\mathrm{RU},n_0} \right) ^2 \label{JC1_d}\\
            &[\mathbf{\bar{J}}_{\mathrm{C}1}]_{2,k}=[\mathbf{\bar{J}}_{\mathrm{C}1}]_{k,2}=\sum_{t_1=1}^{N_{\mathrm{B}}}{\Re \left\{ \frac{8\pi ^2\alpha ^2f_{n_0}^{2}|x_{n_0}|^2}{c^2\sigma ^2}\left( \boldsymbol{\gamma }_{t_1\mathrm{R},n_0}^{\top}\boldsymbol{\gamma }_{\mathrm{RU},n_0} \right) \left( \boldsymbol{\gamma }_{t_1\mathrm{R},n_0}\circledast \boldsymbol{\gamma }_{\mathrm{RU},n_0} \right) ^{\top}\frac{\partial \mathbf{d}_{\mathrm{RU}}}{\partial [\bar{\mathbf{\Theta}}]_k} \right\}} \nonumber
            \\
            &\mspace{180mu}       +\Re \biggl\{ \frac{j\sqrt{2P_{\mathrm{t}}}\alpha ^2|x_{n_0}|^2}{\sigma ^2}\left( \boldsymbol{\gamma }_{t_1\mathrm{R},n_0}^{\top}\boldsymbol{\gamma }_{\mathrm{RU},n_0} \right) \left( \boldsymbol{\gamma }_{t_1\mathrm{R},n_0}\circledast \dot{\mathbf{d}}_{\mathrm{RU}} \right) ^{\top}\frac{\partial \mathbf{d}_{\mathrm{RU}}}{\partial [\bar{\mathbf{\Theta}}]_k} \biggr\}  \nonumber \\
            &\mspace{135mu}\approx \frac{8\pi ^2\alpha ^2f_{n_0}^{2}|x_{n_0}|^2}{c^2\sigma ^2}N_{\mathrm{B}}\left( \boldsymbol{\gamma }_{\mathrm{BR},n_0}^{\top}\boldsymbol{\gamma }_{\mathrm{RU},n_0} \right) \left( \boldsymbol{\gamma }_{\mathrm{BR},n_0}\circledast \dot{\mathbf{d}}_{\mathrm{RU}} \right) ^{\top}\frac{\partial \mathbf{d}_{\mathrm{RU}}}{\partial [\bar{\mathbf{\Theta}}]_k}, \label{JC1_e}\\
            &[\mathbf{\bar{J}}_{\mathrm{C}1}]_{k,l}=\sum_{t_1=1}^{N_{\mathrm{B}}}{\frac{8\pi ^2\alpha ^2f_{n_0}^{2}|x_{n_0}|^2}{c^2\sigma ^2}\left( \boldsymbol{\gamma }_{t_1\mathrm{R},n_0}\circledast \boldsymbol{\gamma }_{\mathrm{RU},n_0} \right) ^{\top}\frac{\partial \mathbf{d}_{\mathrm{RU}}}{\partial [\bar{\mathbf{\Theta}}]_k}\left( \boldsymbol{\gamma }_{t_1\mathrm{R},n_0}\circledast \boldsymbol{\gamma }_{\mathrm{RU},n_0} \right) ^{\top}\frac{\partial \mathbf{d}_{\mathrm{RU}}}{\partial [\bar{\mathbf{\Theta}}]_l}} \nonumber
            \\
            &\mspace{100mu}\,\,      +\Re \left\{ \frac{j2\sqrt{2P_{\mathrm{t}}}\alpha ^2|x_n|^2}{\sigma ^2}\left( \boldsymbol{\gamma }_{t_1\mathrm{R},n_0}\circledast \boldsymbol{\gamma }_{\mathrm{RU},n_0} \right) ^{\top}\frac{\partial \mathbf{d}_{\mathrm{RU}}}{\partial [\bar{\mathbf{\Theta}}]_k}\left( \boldsymbol{\gamma }_{t_1\mathrm{R},n_0}\circledast \dot{\mathbf{d}}_{\mathrm{RU}} \right) ^{\top}\frac{\partial \mathbf{d}_{\mathrm{RU}}}{\partial [\bar{\mathbf{\Theta}}]_l} \right\} \nonumber
            \\
            &\mspace{100mu}\,\,      +\frac{\alpha ^2P_{\mathrm{t}}\lambda _{n}^{2}|x_n|^2}{4\pi ^2\sigma ^2}\left( \boldsymbol{\gamma }_{t_1\mathrm{R},n_0}\circledast \dot{\mathbf{d}}_{\mathrm{RU}} \right) ^{\top}\frac{\partial \mathbf{d}_{\mathrm{RU}}}{\partial [\bar{\mathbf{\Theta}}]_k}\left( \boldsymbol{\gamma }_{t_1\mathrm{R},n_0}\circledast \dot{\mathbf{d}}_{\mathrm{RU}} \right) ^{\top}\frac{\partial \mathbf{d}_{\mathrm{RU}}}{\partial [\bar{\mathbf{\Theta}}]_l},  \nonumber \\
            &\mspace{60mu} \approx \frac{8\pi ^2\alpha ^2f_{n_0}^{2}|x_{n_0}|^2}{c^2\sigma ^2}N_{\mathrm{B}}\left( \boldsymbol{\gamma }_{\mathrm{BR},n_0}\circledast \boldsymbol{\gamma }_{\mathrm{RU},n_0} \right) ^{\top}\frac{\partial \mathbf{d}_{\mathrm{RU}}}{\partial [\bar{\mathbf{\Theta}}]_k}\left( \boldsymbol{\gamma }_{\mathrm{BR},n_0}\circledast \boldsymbol{\gamma }_{\mathrm{RU},n_0} \right) ^{\top}\frac{\partial \mathbf{d}_{\mathrm{RU}}}{\partial [\bar{\mathbf{\Theta}}]_l} \nonumber
            \\
            &\mspace{80mu}\,\, +\frac{\alpha ^2P_{\mathrm{t}}\lambda _{n}^{2}|x_n|^2}{4\pi ^2\sigma ^2}N_{\mathrm{B}}\left( \boldsymbol{\gamma }_{\mathrm{BR},n_0}\circledast \dot{\mathbf{d}}_{\mathrm{RU}} \right) ^{\top}\frac{\partial \mathbf{d}_{\mathrm{RU}}}{\partial [\bar{\mathbf{\Theta}}]_k}\left( \boldsymbol{\gamma }_{\mathrm{BR},n_0}\circledast \dot{\mathbf{d}}_{\mathrm{RU}} \right) ^{\top}\frac{\partial \mathbf{d}_{\mathrm{RU}}}{\partial [\bar{\mathbf{\Theta}}]_l}\label{JC1_f}
        \end{align}
    \end{subequations}
    where $k,l\in\{3,4,5\}$, $\left[ \boldsymbol{\gamma }_{\mathrm{BR},n_0} \right] _r=\frac{\lambda _n}{4\pi d_{\mathrm{B}r}}$, $\forall r$, $d_{\mathrm{B}r}\triangleq \left\| \mathbf{p}_{\mathrm{B}}-\mathbf{p}_r \right\| $.  (\refeq{JC1_c}), (\refeq{JC1_e}), (\refeq{JC1_f}) holds because $\frac{\partial d_{r\mathrm{U}}}{\partial d_{\mathrm{RU}}}$, $\frac{\partial d_{r\mathrm{U}}}{\partial \varphi _{\mathrm{RU}}}$ and $\frac{\partial d_{r\mathrm{U}}}{\partial \theta _{\mathrm{RU}}}$ are all real, and $\boldsymbol{\gamma }_{t_1\mathrm{R},n_0}\approx\boldsymbol{\gamma }_{\mathrm{BR},n_0}, \forall t_1$. This approximation holds because of the assumption that $L_\mathrm{B}\ll d_\mathrm{BR}$.\footnote{\textcolor{blue}{The amplitude differences across the antenna array are far less significant compared with the phase differences. For example, at $f_c=28$GHz, when $d_\mathrm{BR}=3$ m and $N_\mathrm{B}=10\times10=100$ (in this case $L_\mathrm{B}=7.58$ cm), the amplitude variation across the BS array is less than 0.05\%. In fact, as will be shown in Section \ref{sec:numerical results}, the large-scale BS array is not necessary to achieve desirable performance since the COA information is directly sensed by the RIS (passively), not the BS.}} Define $\mathbf{G}_{\mathrm{C}1}\triangleq [\mathbf{\bar{J}}_{\mathrm{C}1}]_{3:5,1:2}[\mathbf{\bar{J}}_{\mathrm{C}1}]_{1:2,1:2}^{-1}[\mathbf{\bar{J}}_{\mathrm{C}1}]_{1:2,3:5}\in \mathbb{R}^{3\times 3}$. Combining (\refeq{JC1_12}), each element of $\mathbf{G}_{\mathrm{C}1}$ can be expressed as
    \begin{align}
        [\mathbf{G}_{\mathrm{C}1}]_{k-2,l-2}=[\mathbf{\bar{J}}_{\mathrm{C}1}]_{1,k}\cdot [\mathbf{\bar{J}}_{\mathrm{C}1}]_{1,l} \left/ [\mathbf{\bar{J}}_{\mathrm{C}1}]_{1,1} \right.+[\mathbf{\bar{J}}_{\mathrm{C}1}]_{2,k}\cdot [\mathbf{\bar{J}}_{\mathrm{C}1}]_{2,l} \left/ [\mathbf{\bar{J}}_{\mathrm{C}1}]_{2,2} \right. ,
    \end{align}
    where $k,l\in\{3,4,5\}$. Combining (\refeq{JC1}) it can be verified that $\mathbf{G}_{\mathrm{C}1}\approx [\mathbf{\bar{J}}_{\mathrm{C}1}]_{3:5,3:5}$. Therefore, in Case 1 we have $\mathbf{\bar{J}}_{\mathrm{E}}(\boldsymbol{\eta })\approx\mathbf{0}$. In Case 2, since the bandwidth of the signal $f_N-f_c\ll f_c$, therefore we also have $\boldsymbol{\gamma }_{b_0\mathrm{R},t_2}\approx \boldsymbol{\gamma }_{b_0\mathrm{R},0}, \forall t_2$, where $\left[ \boldsymbol{\gamma }_{b_0\mathrm{R},0} \right] _r=\frac{\lambda _c}{4\pi d_{b_0r}}, \forall r$.  Through a similar procedure, it can be verified that the same result holds in Case 2, and the detailed proof is omitted.
}

\end{document}